\documentclass{article}

    \usepackage[final]{neurips_2024}

\usepackage[dvipsnames]{xcolor}
\usepackage[utf8]{inputenc} 
\usepackage[T1]{fontenc}    
\usepackage{hyperref}       
\usepackage{url}            
\usepackage{booktabs}       
\usepackage{amsfonts}       
\usepackage{nicefrac}       
\usepackage{microtype}      
\usepackage{xcolor}         
\usepackage{lipsum}
\usepackage{graphicx}
\usepackage{amsmath}
\usepackage{titlesec}
\usepackage{tabularx}
\usepackage{booktabs}
\usepackage{multirow} 
\usepackage{amssymb}  
\usepackage{graphicx}  
\usepackage{booktabs} 
\usepackage{tabularx}  
\usepackage{ragged2e}  
\usepackage{caption}
\usepackage{algorithm}
\usepackage{algorithmic}

\usepackage{tcolorbox}
\usepackage{wrapfig}
\usepackage{colortbl}

\usepackage{amsmath}

\definecolor{codegray}{rgb}{0.5,0.5,0.5}
\definecolor{codepurple}{rgb}{0.58,0,0.82}
\definecolor{backcolour}{rgb}{0.95,0.95,0.92}
\definecolor{deepgreen}{rgb}{0.0, 0.5, 0.0}

\usepackage{booktabs} 
\usepackage{makecell} 

\usepackage{caption} 

\usepackage{colortbl}
\definecolor{customblue}{HTML}{74AED4} 
\definecolor{customgreen}{HTML}{D3E2B7} 
\definecolor{customred}{HTML}{ECA8A9} 
\definecolor{custompurple}{HTML}{CFAFD4}
\definecolor{customorange}{HTML}{FFD599}

\definecolor{tableorange}{HTML}{FFD599}
\definecolor{tablepurple}{HTML}{E3D4E7}
\definecolor{tableblue}{HTML}{D9E8FA}
\definecolor{tablegreen}{HTML}{D6E9D6}

\usepackage{float}
\usepackage{xspace}
\definecolor{darkgreen}{rgb}{0.0, 0.5, 0.0}
\definecolor{darkgray}{gray}{0.4}
\definecolor{maroon}{rgb}{0.5, 0.0, 0.0}
\definecolor{navy}{rgb}{0.0, 0.0, 0.5}
\definecolor{teal}{rgb}{0.0, 0.5, 0.5}
        \definecolor{lightblue}{rgb}{0.85,0.92,0.97}

\usepackage{wrapfig}

\usepackage{pifont} 
\usepackage{textcomp} 
\usepackage{booktabs}
\tcbuselibrary{breakable}

\usepackage{tikz}

\usepackage{amsthm}
\theoremstyle{plain}

\title{\Large WaveMind: Towards a Conversational EEG Foundation Model Aligned to Textual and Visual Modalities}

\author{
  \textbf{Ziyi Zeng$^1$},
  \textbf{Zhenyang Cai$^1$},
  \textbf{Yixi Cai$^1$},
  \textbf{Xidong Wang$^1$},
  \textbf{Junying Chen$^1$},
  \textbf{Rongsheng Wang$^1$},\\
  \textbf{Yipeng Liu$^1$},
  \textbf{Siqi Cai$^2$},
  \textbf{Benyou Wang$^1$}\thanks{Benyou is the corresponding author.}~,
  \textbf{Zhiguo Zhang$^2$},
  \textbf{Haizhou Li$^1$},
  \\
  $^1$ The Chinese University of Hong Kong, Shenzhen\\
  $^2$ Harbin Institute of Technology, Shenzhen\\
   \texttt{wangbenyou@cuhk.edu.cn}\\ 
}

\newcolumntype{Y}{>{\centering\arraybackslash}X}

\begin{document}

\maketitle

\begin{abstract}

   Electroencephalography (EEG) interpretation using multimodal large language models (MLLMs) offers a novel approach for analyzing brain signals. However, the complex nature of brain activity introduces critical challenges: EEG signals simultaneously encode both cognitive processes and intrinsic neural states, creating a mismatch in EEG paired-data modality that hinders effective cross-modal representation learning. Through a pivot investigation, we uncover complementary relationships between these modalities. Leveraging this insight, we propose mapping EEG signals and their corresponding modalities into a unified semantic space to achieve generalized interpretation. To fully enable conversational capabilities, we further introduce WaveMind-Instruct-338k, the first cross-task EEG dataset for instruction tuning. The resulting model demonstrates robust classification accuracy while supporting flexible, open-ended conversations across four downstream tasks, thereby offering valuable insights for both neuroscience research and the development of general-purpose EEG models.

\end{abstract}


\section{Introduction}

\begin{wrapfigure}[16]{r}{0.5\textwidth}
  \centering
  \vspace*{-20pt}\includegraphics[width=0.5\textwidth]{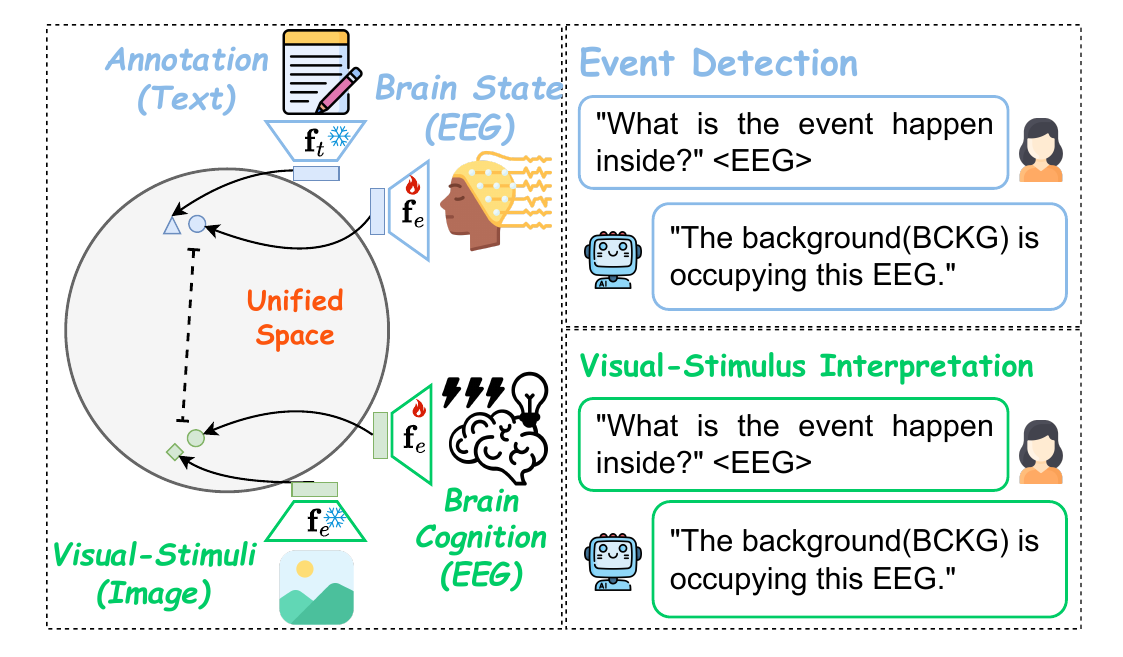}
  \caption{\textbf{Overall illustration of WaveMind with support for downstream tasks} \textit{Left}: The model is compatible with more upstream data through shared space. \textit{Right}: Example of EEG interpretation over various downstream tasks.}
  \label{fig:abstract}
\end{wrapfigure}

Electroencephalography (EEG), a non-invasive technique for recording brain activity, holds significant promise for integration with Multimodal Large Language Models (MLLMs). While MLLMs demonstrate robust capabilities in dialogue and multi-task scenarios, their application to EEG interpretation remains constrained by critical limitations: existing EEG foundation models support diverse downstream tasks but lack conversational proficiency, whereas dedicated EEG conversational models are restricted to single-task execution, with their interactive potential not fully explored.

To enable conversational EEG interpretation, the common practice is to align EEG with paired data modalities in advance, and then hand it over to LLM for natural language generation, which can divide into two category by modality: The \textit{image alignment} paradigm reconstructs external visual stimuli from neural data \cite{liRealMindZeroShotEEGBased2024, mishraThought2TextTextGeneration2024}, while the \textit{text alignment} paradigm correlates neural states with textual annotations such as emotional valence or clinical descriptors \cite{kim2024eeggptexploringcapabilitieslarge, zhou2024belt2bootstrappingeegtolanguagerepresentation, chen2025eegemotioncopilotoptimizing}. We contend that \textbf{previous alignment approaches using only a single paired modality undermine the utility of training data sources and constrain model generalizability.} However, the research community currently lacks chat-oriented foundation models and open-source EEG instruction-tuning datasets, which are the focus of our work.

Our pivot investigation reveals that different paired modalities originating from distinct sources are complementary. To further advance this approach, we scaled up and preprocessed data from five diverse datasets, and constructed \textit{WaveMind-Instruct}, the first open-source EEG instruction-tuning dataset. Through this process, we developed \textit{WaveMind}, a cross-task EEG foundation model that exhibits strong EEG-awareness and conversational capabilities. Our contributions are listed below:

\begin{itemize}
    \item We introduce a \textbf{multimodal EEG alignment framework}, which aligns EEG with textual and visual modalities. This approach broadens the scope of upstream data and facilitates downstream interpretation without architectural modifications.

    \item We introduce \textbf{WaveMind-Instruct}, an open-source EEG instruction-following dataset featuring three instruction types and two dialogue scenarios. Upon this, we develop \textbf{WaveMind}, the first conversational EEG foundation model for unified interpretation across diverse brain activities.
\end{itemize}

\section{Related Work}
    \label{sec:related_work}

    \paragraph{Representation Alignment}

    Representation Alignment has emerged as a pivotal paradigm for bridging neural signals with brain activity. NICE \cite{songDecodingNaturalImages2024} pioneered the alignment of EEG representations with image features to achieve EEG-based visual object classification, validating the feasibility of mapping neural activity to high-level visual semantics. Subsequently, RealMind \cite{liRealMindZeroShotEEGBased2024} and Thought2Text \cite{mishraThought2TextTextGeneration2024} successfully enabled high-fidelity language decoding from EEG signals. Consequently, representation alignment becomes an important method for EEG-to-text decoding.




    \paragraph{EEG-to-Text Decoding}

    There are two approaches in EEG-to-Text Decoding: (1) EEG-to-Text \textit{Translation}, which aims to accurately reconstruct thought in word-level text\cite{duanDeWaveDiscreteEEG2024,wangOpenVocabularyElectroencephalographyToText2024,kostas2021bendrusingtransformerscontrastive}. However, this requires massive fine-grained annotations at the millisecond level and experiences poor performance when across languages and individuals. (2) EEG-to-text \textit{Interpretation}, which interprets EEG from a higher-dimensional perspective and aims to achieve semantic accuracy\cite{liRealMindZeroShotEEGBased2024,mishraThought2TextTextGeneration2024,kim2024eeggptexploringcapabilitieslarge,chen2025eegemotioncopilotoptimizing} . This only requires coarse-grained information paired with EEG, and makes it possible to extend to a wider range of scenarios and tasks. Our work (WaveMind) is the pioneering work in this latter approach. We listed and compared our work with other EEG interpretation studies in Table \ref{tab:previous_work}.

     \begin{table*}[h!]
    \scriptsize
    \resizebox{\columnwidth}{!}{
    \begin{tabular}{p{2.5cm} c c c cccccc c} 
    \toprule
    \multirow{3}{*}{\textbf{Paper}} & 
    \multirow{3}{*}{\textbf{\makecell{Chatbot\\Like}}} &
    \multirow{3}{*}{\textbf{\makecell{EEG\\Awareness}}} & 
    \multirow{3}{*}{\textbf{\makecell{Paird\\Modality}}} &
    \multicolumn{1}{c}{\textbf{Brain Cognition}} & \multicolumn{3}{c}{\textbf{Brain State}} &
    \multirow{3}{*}{\textbf{\makecell{Instruction\\Tuning}}} \\
    \cmidrule(lr){5-5}  \cmidrule(lr){6-8} 
    & & & & \multicolumn{1}{c}{\makecell{Visual-Stimulus\\Interpretation}\includegraphics[scale=0.02]{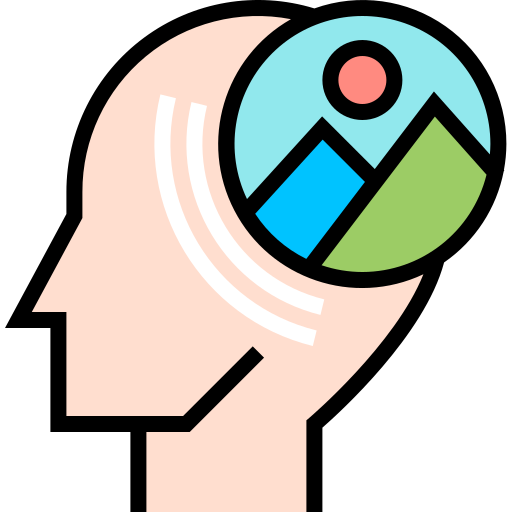}} 
    & \multicolumn{1}{c}{\makecell{Emotion\\Recognition}\includegraphics[scale=0.03]{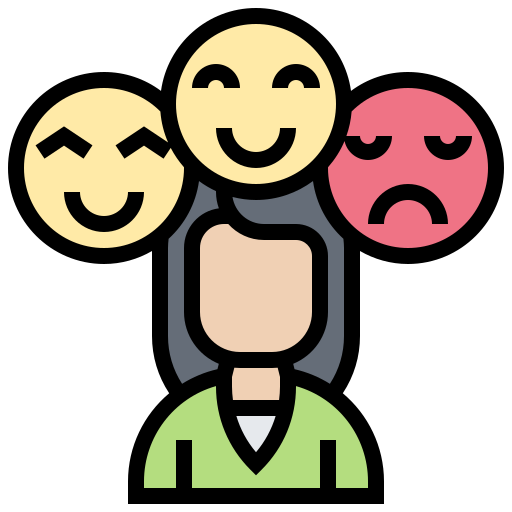}} 
    & \multicolumn{1}{c}{\makecell{Event\\Detection}\includegraphics[scale=0.03]{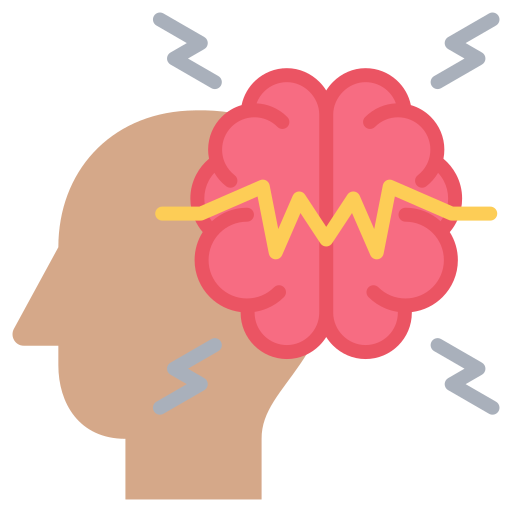}} 
    & \multicolumn{1}{c}{\makecell{Abnormality\\Detection}\includegraphics[scale=0.03]{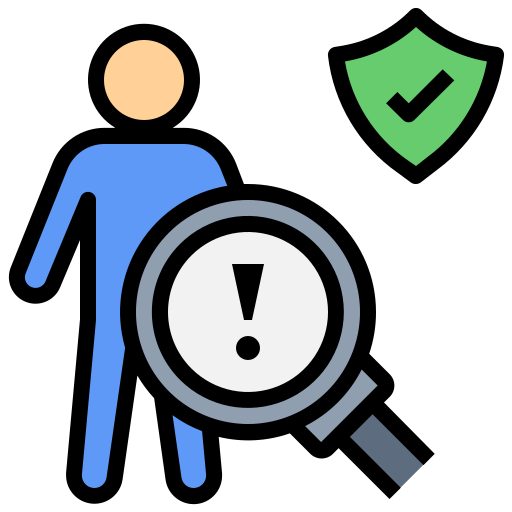}} 
    & \\
    \midrule
    \multicolumn{4}{l}{\textbf{Image Alignment}} & & & & & \\ 
    Real Mind\cite{liRealMindZeroShotEEGBased2024} & \ding{51} & \ding{55} & Image & \ding{51} & \ding{55} & \ding{55} & \ding{55} & \ding{55} \\ 
    Thought2Text\cite{mishraThought2TextTextGeneration2024}\textsuperscript{*} & \ding{51} & \ding{55} & Image & \ding{51} & \ding{55} & \ding{55} & \ding{55} & \ding{51} \\
    NICE++\cite{mishraThought2TextTextGeneration2024}\textsuperscript{*} & \ding{51} & \ding{55} & Image & \ding{51} & \ding{55} & \ding{55} & \ding{55} & \ding{51} \\
    
    \midrule
    \multicolumn{4}{l}{\textbf{Text Alignment}} & & & & & \\ 
    EEG-GPT\cite{kim2024eeggptexploringcapabilitieslarge} & \ding{51} & \ding{51} & Text & \ding{55} & \ding{55} & \ding{55} & \ding{51} & \ding{55} \\ 
    BELT-2\cite{zhou2024belt2bootstrappingeegtolanguagerepresentation} & \ding{55} & \ding{51} & Text & \ding{55} & \ding{51} & \ding{55} & \ding{55} & \ding{51} \\
    
    Brain Emotion Copilot\cite{chen2025eegemotioncopilotoptimizing} & \ding{51} & \ding{51} & Text & \ding{55} & \ding{51} & \ding{55} & \ding{55} & \ding{51} \\

    ARIEL\cite{inproceedings} & \ding{51} & \ding{51} & Text & \ding{55} & \ding{51} & \ding{55} & \ding{55} & \ding{55} \\

    
    \midrule
    \textbf{Ours\textsuperscript{*†}} & \ding{51} & \ding{51} & Both & \ding{51} & \ding{51} & \ding{51} & \ding{51} & \ding{51} \\ 
    
    \bottomrule
    \end{tabular}
    }

    \begin{minipage}{1\textwidth}
    {\tiny
        \textsuperscript{*} represents code open-source,
        \textsuperscript{†} represents checkpoint open-source
    }
    \end{minipage}
    
    \caption{\textbf{Related Works Comparison of EEG Interpretation using EEG-MLLM.} Only chatbot-like EEG-MLLMs are included. }
    \label{tab:previous_work}
    \end{table*}

    \paragraph{EEG Foundation Models}

    EEG Foundation Models (EEG-FMs) refer to large-scale, generic models trained on diverse datasets to enable cross-task generalization in brain signal analysis. Early explorations in this domain have yielded models such as Brant \cite{yuan2024brainwavebrainsignalfoundation}, BIOT \cite{yang2023biot}, and EEGPT \cite{yue2024eegptunleashingpotentialeeg}, which demonstrate strong cross-task performance. More recently, the emergence of EEG-MLLMs, along with works like NeuroGPT \cite{cui2024neurogptfoundationmodeleeg}, BrainBERT \cite{wang2023brainbertselfsupervisedrepresentationlearning}, NeuroLM \cite{jiang2025neurolmuniversalmultitaskfoundation}, and BrainOmni \cite{xiao2025brainomnibrainfoundationmodel}, shows significant promise in advancing probabilistic modeling of neural data. However, a critical limitation persists: current EEG-FMs focus narrowly on classification tasks. Moreover, These models lack conversational capabilities and fail to leverage Brain Cognition data, which captures higher-order cognitive processes like language comprehension.



\section{Pilot Study on Fusion of Multiple Paired Modalities}

    

    \paragraph{Multiple Paired Modalities}
    Due to the diversity of experimental paradigms, the paired data modalities recorded concurrently with EEG are highly heterogeneous. \textit{Images} and \textit{text} are common paird modalities. Images are used as visual stimuli to activate cognitive-related brain activity in participants, which we call \textit{Brain Cognition} \cite{gaoSpecificEndophenotypesEEG2024}. Text usually originates from external annotations and is used to describe the internal state of the brain in resting-state \cite{baiReviewRestingStateElectroencephalography2017}, which we call the \textit{Brain State}. This poses substantial challenges for comprehensive brain modeling.

    
    
\begin{figure}[h]
        \centering
        \includegraphics[width=0.8\textwidth]{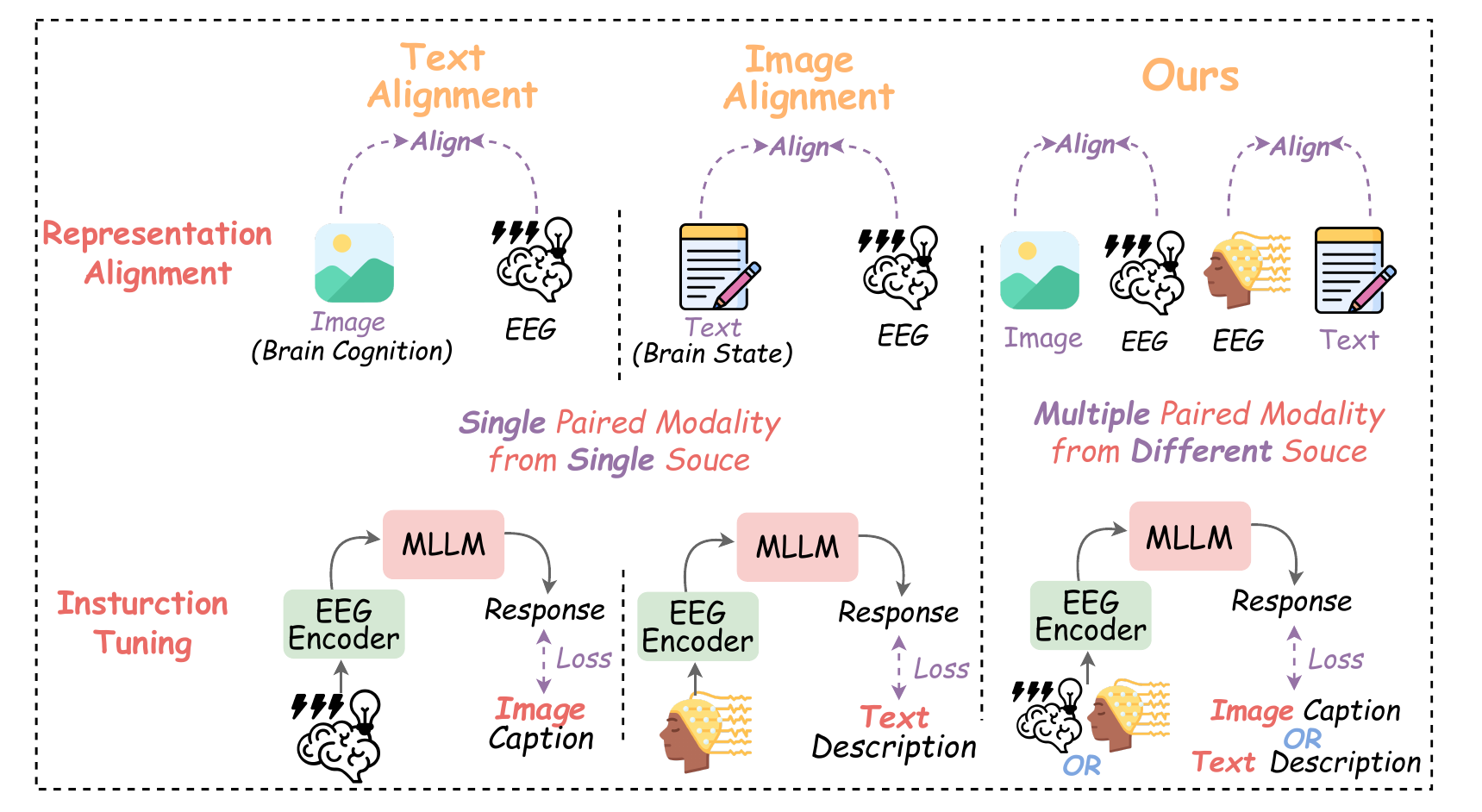}
        \caption{\textbf{The Comparison of Two Typical Alignment Methods with Ours.} The proposed method can effectively enhance the adaptability of upstream data and the generalization of downstream tasks.}
        \label{fig:AlignmentExample}
    \end{figure}

    As illustrated in Figure \ref{fig:AlignmentExample}, current approaches for natural language interpretation primarily align EEG signals with either images (Image Alignment) or text (Text Alignment), yet exclusively rely on single-source training data, which significantly constrains their capacity to leverage diverse upstream resources. To date, no existing work has investigated aligning EEG with multiple heterogeneous multimodal sources derived from distinct origins.

    \paragraph{Pilot Study} 
    
     We assume that combining multiple pair modalities from different sources is beneficial for alignment and language generation. However, it requires that the presentations of both paired modalities are already in the same space before alignment, which the CLIP model \cite{radford2021learningtransferablevisualmodels} provides an ideal solution. Therefore, we selected two datasets of THING-EEG and TUEV, which represent Brain Cognition (Image-EEG) and Brain State (Text-EEG), respectively. First, we perform alignment on each paired modality (dataset) separately. Then, we combined both datasets to train the EEG encoder jointly, simulating the integration of multiple paired modalities.

    \begin{table}[hbtp]
        \centering
        \small
        \renewcommand{\arraystretch}{1.2}
        \begin{tabularx}{0.9\textwidth}{l>{\centering\arraybackslash}X>{\centering\arraybackslash}X} 
        \toprule
          & \textbf{Brain Cognition (Image)\includegraphics[scale=0.03]{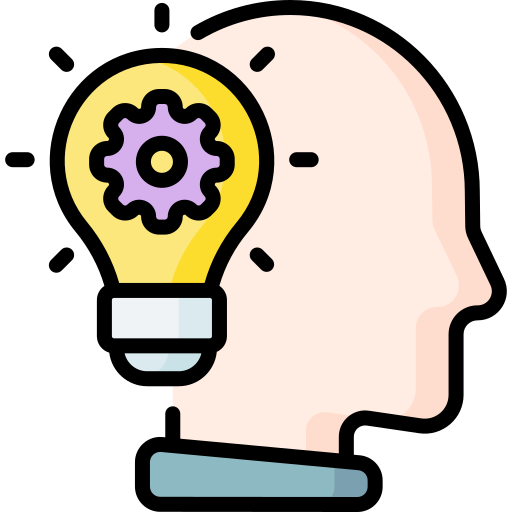}}  & \textbf{Brain State (Text)\reflectbox{\includegraphics[scale=0.033]{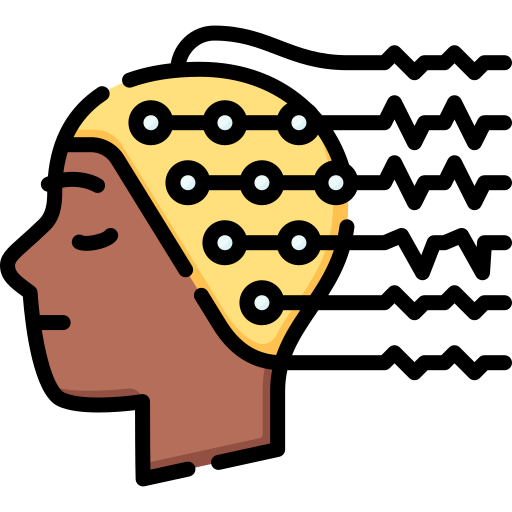}}} \\ 
        \midrule
        \textbf{Image-EEG Only\textsuperscript{\dag}} & 0.648 & --\textsuperscript{*} \\  
    
            \textbf{Text-EEG Only\textsuperscript{\ddag}} & -- & 0.742 \\ 
        \midrule
         \textbf{Fusion of Both Modalities} & \textbf{0.671} \textcolor{red}{\scriptsize(+0.023)} & \textbf{0.788} \textcolor{red}{\scriptsize(+0.046)} \\ 
        \bottomrule
        \end{tabularx}
        
        \begin{minipage}{0.9\textwidth}
        {\scriptsize
            \textsuperscript{\dag} The 2-way retrieval accuracy using \textit{THING-EEG-8K} is reported. \\
            \textsuperscript{\ddag} The 6-way retrieval accuracy using \textit{TUEV-8k} is reported. \\
            \textsuperscript{*} -- indicates not applicable.
        }
        \end{minipage}
    
        \caption{\textbf{Feature Retrieval Accuracy with Brain Cognition and Brain State.} The EEG Encoder is trained to align with CLIP space. }
        \label{tab:pivot_study}
    \end{table}

    \paragraph{Analysis} As shown in Table~\ref{tab:pivot_study}, models trained on a single paired modality achieved a certain level of EEG understanding, but they all lagged behind the performance of the model trained with multiple paired modalities. This suggests that \textbf{there is a degree of generalization between different paired modalities and sources}, which is worth further exploration. 
    Moreover, from the perspective of neuroscience, Brain Cognition and Brain State, as two expressions of brain activity, demonstrate complementarity. 



\section{Data Engineering}
\label{sec:data_eng_}
    
    After identifying that integrating multiple paired modalities improves model understanding, we expanded our scope to a larger scale, gathering an extensive dataset to exploit this finding further.


    \subsection{Data Sources}
    
        \paragraph{Brain Cognition (Image-EEG) Datasets}  We include 2 datasets with each distinct object set: 
        \texttt{THING-EEG} \cite{giffordLargeRichEEG2021,grootswagersHumanEEGRecordings2022} contains 1573 object categories, supporting zero-shot testing for 200 classes.  
        \texttt{ImageNet-EEG} \cite{palazzoGenerativeAdversarialNetworks2017,palazzoDecodingBrainRepresentations2021} comprises 40 object categories. Among them, each EEG is paired with a corresponding visual-stimulus image.  
        
        \paragraph{Brain State (Text-EEG) Datasets}   We include 3 Brain State datasets of Text-EEG:
        \texttt{TUEV} \cite{obeidTempleUniversityHospital2016} provides six-class event-related annotations in EEG.  
        \texttt{TUAB} \cite{obeidTempleUniversityHospital2016} focuses on binary labeling of EEG abnormalities.  
        \texttt{SEED} \cite{zheng2015investigating,duan2013differential} annotates EEG data with emotional state.  
        Among them, each EEG is paired with a corresponding doctor's annotation in text-based.
        

        \subsection{EEG Preprocessing}

        We implemented a unified preprocessing pipeline across all EEG datasets. Raw signals were first aligned to pre-defined 32 channels in 10-20 system and resampled to 512 Hz. Subsequently, data segments were standardized to fixed 1-second durations through replication of shorter segments or partitioning of longer ones. Further preprocessing details are provided in Appendix \ref{sec:data_preprocerssing}. The final preprocessed data were structured as spatiotemporal tensors in \(\mathbb{R}^{32 \times 512}\).


        \begin{figure*}[t]
            \centering
            \includegraphics[width=1\linewidth]{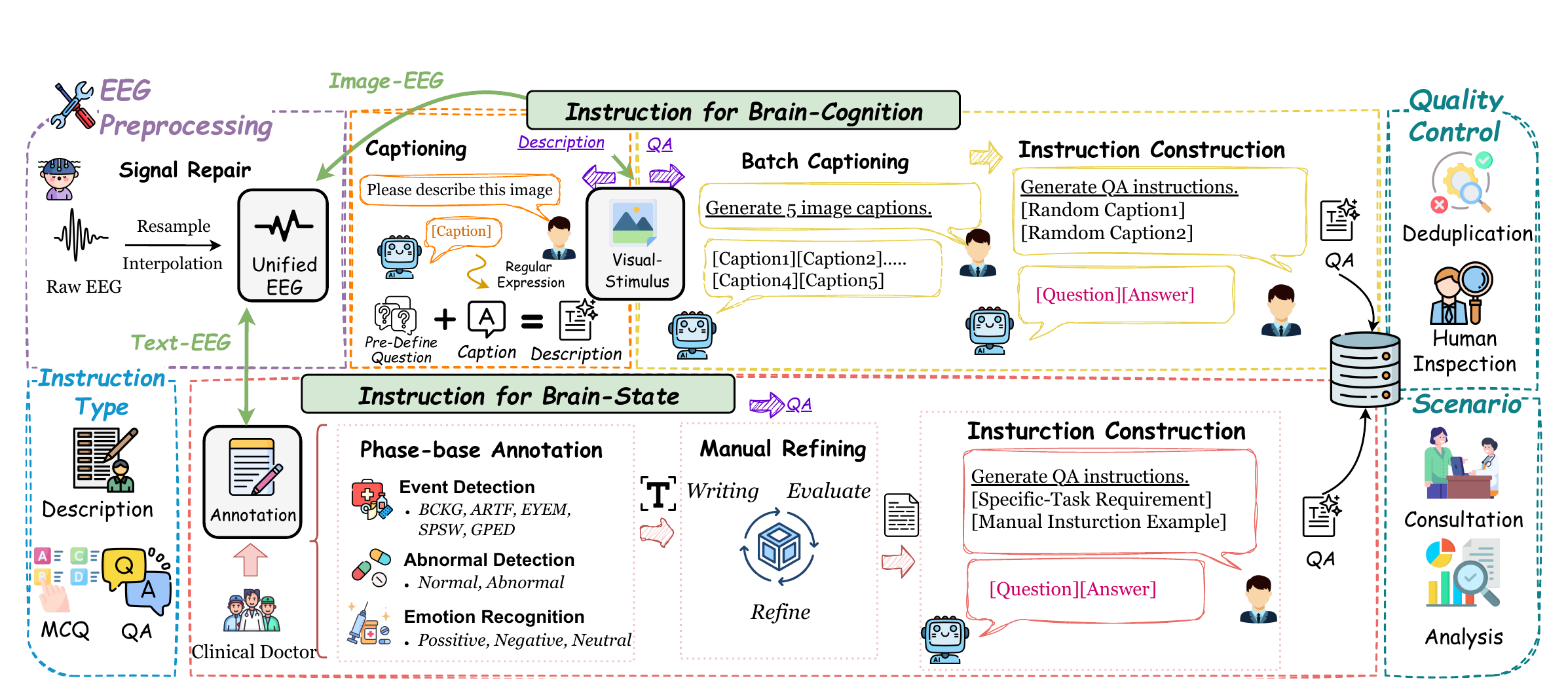}
            \caption{\textbf{Instruction Construction Pipeline of WaveMind.} The raw EEG signals are first pre-processed into segments with the same configuration, then executed with different instruction construction processes depending on the type of labels. We have constructed four types of instructions to ensure the model learns diverse knowledge.}
            \label{fig:trainingStage}
        \end{figure*}


        \subsection{Training Data} 


        \subsection{Training Data Statistics}

       \begin{table}[tphb]
            \centering
            \footnotesize
            \begin{tabular}{m{2.4cm} >{\centering\arraybackslash}c >{\centering\arraybackslash}c >{\centering\arraybackslash}c >{\centering\arraybackslash}l
            }
                \toprule
                \textbf{Dataset} & \multicolumn{3}{c}{\textbf{Training Stage}} & \textbf{Downstream Task} \\
                \cmidrule{2-4}
                 & \textbf{I} & \textbf{II} & \textbf{III} & \\
                \midrule
                LLAVA\_Pretrian & - & 558K & - & - \\
                THING-EEG & 528K & - & 153K & Visual-Stimulus Interpretation  \\
                ImageNet-EEG & 8K & - & 103K & Visual-Stimulus Interpretation \\
                SEED & 122K & - & 28K & Emotion Recognition \\
                TUEV & 113K & - & 28K & Event Detection \\
                TUAB & 548K & - & 39K & Abnormality Detection \\
                \midrule
                \textbf{Total} & 1.3M & 558K & 338K & - \\
                \bottomrule
            \end{tabular}
            \caption{
                \textbf{Training Data Statistics. }
            }
            \label{tab:dataStatistic}
        \end{table}

        
        The training data statistics are presented in Table \ref{tab:dataStatistic}. During Encoder Alignment (Stage I), we utilized the original data, yielding 1.3M samples for representation alignment (Session \ref{sec:Data_alignment}). During Cold-Start (Stage II), we use \textit{LLAVA\_Pretrain} to initialize the modality adapter. During Instruction Tuning (Stage III), we applied a random sampling strategy to ensure sample balance across datasets (Session \ref{sec:Data_Instruction}). This yields \textit{WaveMind-Instruct}, which generates 338K samples in total. To ensure data balance, we conducted repeated sampling on datasets with limited data. All datasets undergo the same preprocessing pipeline, which can be found in the Appendix \ref{Detail Data Statistic}.
    
        \subsubsection{Training Data for Encoder Representation Alignment}
        \label{sec:Data_alignment}




        For image pair modality (Brain Cognition), we directly adopt the image feature $\mathbf{\hat{Z}}_I$ as alignment targets to avoid losing information for Image-EEG data. However, for text pair modality (Brain State), only phase-level category annotations are available. We argue that such coarse-grained information is insufficient to support effective representation alignment. Therefore, we manually construct fine-grained descriptions for each annotation, which can be found in Appendix \ref{sec:Hand-written_Description_appendix}.


    \subsubsection{Training Data for Instruction Tuning}
    \label{sec:Data_Instruction}
    The \textit{WaveMind-Instruct} combines three instruction formats in its training data: description, question-answer pairs (QA), and multiple-choice questions (MCQ). For MCQ instruction, the number of options matches the actual class count in each dataset. For the remaining instruction type, we employ customized construction pipelines, as shown in Figure \ref{fig:trainingStage}. Detailed implementation is provided in Appendix \ref{sec:synthesis_process}.

        \paragraph{Instruction Construction for Brain Cognition (Image-EEG)} 
        The modality of pair data to Brain Cognition is image. Therefore, to build the \textit{description} instructions, we firstly ask the Visual-MLLM (e.g., Qwen2.5-VL\footnote{We manually checked several Visual-MLLMs including Qwen2.5-VL, LLAVA1.5, and DeepSeeke V3. Qwen2.5-VL achieved Competitive results with a lower price. Therefore, Qwen2.5-VL was used in this study to generate image captions.}) to generate image captions. Then we used regular expressions to remove image-related keywords and replaced them with EEG-related expressions. To build the \textit{QA} instruction, we ask LLM (e.g., Qwen2.5-Instruct) to transform image captions into diverse QA, which are based on closed-ended and open-ended.
        
        \paragraph{Instruction Construction for Brain State (Text-EEG) } 
        The modality of pair data to the Brain State is text. To generate diverse QA instructions, we first manually collect extensive factual definitions relevant to each annotation. Then, we compose and refine these facts into QA instructions to serve as a seed. Finally, we leverage LLM to rewrite these QA instructions, matching their tone and imitating their semantics to ensure diversity in the synthesized instructions.

        \paragraph{Instruction Scenario} We designed two scenarios for distinct downstream tasks: the \textit{consultation} scenario, where WaveMind and the human assume the roles of doctor and patient, specifically covering event detection and abnormality detection due to their relevance to disease diagnosis; and the \textit{analysis} (non-medical) scenario, where the specific roles of WaveMind and the human remain undefined.

       \paragraph{Instruction Quality Control}
       To mitigate hallucinations, we first manually inspected the raw inputs to guide LLM in the instruction synthesis process, which are image captions and factual definitions, respectively. The averaged acceptance rates from 3 invited reviewers reached an average score of 0.945 and 0.966 for captions and definitions. We then proceeded to check the instructions generated by the rejected raw input and found that they were still acceptable. Additionally, we explicitly excluded EEG-related details during synthesis. To enhance instruction diversity, we applied 2-gram and ROUGE-L deduplication to ensure diversity, retaining approximately 70\% of the synthesized data in the final dataset. Manual inspection results and more details can be found in Appendix \ref{sec:human_eval}.

\section{WaveMind}
   In this section, we will introduce the model architecture by module and our proposed training stage based on a unified latent space.

  \subsection{Architecture}

\begin{figure*}[t!]
    \centering
    \includegraphics[width=1\linewidth]{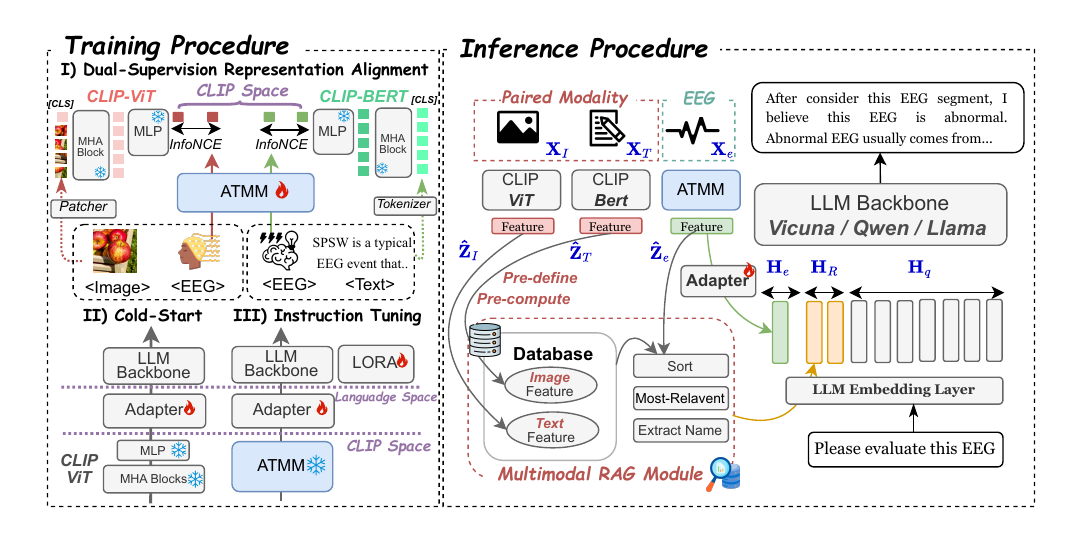}
    \caption{\textbf{The Overall Architecture of WaveMind.} Left: three-stage training procedure. Right: inference procedure of WaveMind. The system projects EEG data into a unified semantic space and integrates retrieval-augmented generation (RAG) for robust language generation.}
    \label{fig:architecture_WaveMind}
    \end{figure*}

    Given raw EEG signals $X_e \in \mathbb{R}^{T \times C}$, we aim to decode them into natural language tokens $W = \{w_1, \dots, w_N\}$ via an LLM backbone, as illustrated in Figure \ref{fig:architecture_WaveMind}. For the EEG encoder, we proposed ATMM as an enlarged version of ATM-S \cite{liVisualDecodingReconstruction2024} to meet the requirements of massive training data. It processes $X_e$ to obtain EEG feature $Z_e$, which undergoes $L_2$-normalization to produce $\hat{Z_e}$. This normalized representation is transformed via a modality adapter into $H_e \in \mathbb{R}^{1 \times d_h}$. For language instructions $X_q$, the token embedding layer of LLM backbone generates  $H_q \in \mathbb{R}^{l_q \times d_h}$ where $l_q$ denotes sequence length.

    Many works show that inserting category information is beneficial for model generation \cite{baiDreamDiffusionGeneratingHighQuality2023,mishraThought2TextTextGeneration2024,lanSeeingBrainImage2023}. Therefore, WaveMind incorporates a Retrieval-Augmented Generation (RAG) module that stores multimodal supervision's features (i.e. $\hat{Z}^I$ and $\hat{Z}^T$) with their category. When $\hat{Z_e}$ is provided, the module retrieves the most similar features to get their corresponding category. More details can be found in Appendix E.2. Therefore, the EEG tokens $H_e$, retrieved category token $H_r$, and instruction tokens $H_q$ are concatenated as below:
    \begin{equation*}
    [H_e; H_r; H_q] \in \mathbb{R}^{(1 + l_r + l_q) \times d_h}
    \end{equation*}

    The joint representation is fed into the LLM base and decoded into neutral language.


    \subsection{Training Paradigm}

    \paragraph{Stage I: Encoder Representation Alignment}

    \label{sec:stage1}

    We perform alignment using \textit{clip-vit-large\_patch-14-336} \cite{radford2021learningtransferablevisualmodels}. CLIP-ViT generates visual features $\mathbf{Z}_I \in \mathbb{R}^{768}$ from image supervision, while CLIP-BERT produces semantic features $\mathbf{Z}_T \in \mathbb{R}^{768}$ from text supervision. After L2 normalization to get $\mathbf{\hat{Z}}_I$ and $\mathbf{\hat{Z}}_T$, they both in the same CLIP space. The loss of Stage I can be defined as:
    
    \begin{equation}
        \mathcal{L} = \lambda \mathcal{L}_{\text{img}} + (1-\lambda)\mathcal{L}_{\text{txt}}
    \end{equation}
        
    Where $\mathcal{L}_{\text{img}}$ and $\mathcal{L}_{\text{txt}}$ denote the InfoNCE loss within the Image-EEG pair and the Text-EEG pair, respectively. We then train ATMM over 1.2M EEG pairs with the other 7 baseline EEG encoders for comparison. Additionally, our practice shows that adding the auxiliary loss $\mathcal{L}_{cls}$ based on specific categories showed no performance improvement, as shown in Appendix \ref{sec:loss_vs}. 


    \paragraph{Stage II: Cold-start for CLIP Space Adaptability}

    Given the adapter's critical role in bridging CLIP and language spaces, proper initialization is essential. We implement a cold-start strategy by pre-training the adapter on image-domain data $\mathbf{\hat{Z}}_I$ using the \textit{LLaVA-Pretrain-558K} dataset. This leverages the shared CLIP space between image features ($\mathbf{\hat{Z}}_I$) and EEG features ($\mathbf{\hat{Z}}_e$). This aligns the MLLM with CLIP space and initializes EEG-domain tuning.
    

    \paragraph{Stage III: EEG Instruction Tuning}
    At this stage, we perform instruction tuning using the \textit{WaveMind-Instruct-338K}. In this stage, the LoRA module and modality-adapter are unfrozen during training, while ATMM is frozen during training.


\section{Experiments}

    \subsection{Experimental Setting}
    \subsubsection{Training Setting}
    The LLM backbone is \textit{Vicuna-1.5-7B}, augmented with a two-layer MLP modality adapter, while the EEG encoder architecture is ATMM. In Stage I, the training takes approximately 13 hours per encoder on a single A40 GPU. Stage II and Stage III utilized 4 NVIDIA A800 GPUs and were trained for 15 hours. Additional training settings can be found in Appendix \ref{Training Setting Detail}.

    \subsubsection{WaveMind-Bench}
    

    \begin{wraptable}[11]{r}{0.48\columnwidth}
        \vspace{-2em}
        \begin{tcolorbox}[title=WaveMind-Bench Example]
            \hspace{-1em}{
            \begin{minipage}{0.25\linewidth}
                \centering
                \includegraphics[width=0.8\linewidth]{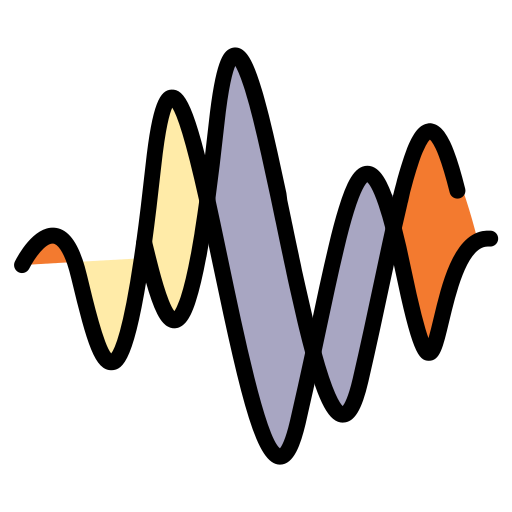} \\
                \textbf{EEG Signal}
            \end{minipage}
            }
            \hfill
            \begin{minipage}{0.72\linewidth}
                What emotional moment is captured in this man's EEG? Choose one letter. \\
                (A) Negative \quad (B) Neutral \quad (C) Positive
            \end{minipage}
        \end{tcolorbox}
        \caption{\textbf{Multiple Choice Question (MCQ) Example in WaveMind-Bench.} Models select answers from \textit{k} options.}
    \end{wraptable}

    There are currently no benchmarks available to test the ability of chat-like EEG-MLLM. Therefore, we sample from the test set and use it to construct MCQs. There are 3 types of options in MCQ: 2, 4, and $k$, where $k$ represents the actual number of classes in the dataset. Specifically, for the THING-EEG dataset, which has an exceptional 200 categories in zero-shot, we set $k$ to 40 to maintain practicality. We refer to it as \textit{WaveMind-Bench-12k}. Due to the strong privacy and limitations of EEG datasets, we will open-source the construction code to enhance reproducibility.

    \subsubsection{Evaluation Metrics}
        This study utilizes weighted K-way accuracy to evaluate the classification ability of WaveMind (Session \ref{WaveMind_Classification_Assessment}) and the perception ability of the encoder (Session \ref{sec:eeg-perception-on-encoder}). In this metric, each question consists of $K$ candidates, including one correct answer and $K-1$ randomly selected incorrect categories. For Natural Language Generation (NLG) evaluation, we adopt a comprehensive set of metrics, comprising three model-free and two model-based measures. The model-free metrics include BLEU-1/2, METEOR, and ROUGE-1/2, which assess lexical and structural overlap. For model-based evaluation, we employ \textit{MiniLm-L12-v2} to measure semantic similarity and GPT-4o to compute the matching score, ensuring a robust assessment of generated text quality.

\subsection{Evaluation Results}

    We evaluate WaveMind's capabilities from MCQ classification (using WaveMind-Bench) to conversational assessment, as reliable EEG interpretation in EEG MLLMs necessitates accurate category identification as a foundation.

    \subsubsection{Classification Assessment}
    \label{WaveMind_Classification_Assessment}
    \paragraph{Subject-Dependent Evaluation}

    We evaluate the MLLM's capacity to identify visual-stimulus objects and annotation categories within EEG signals using \textit{WaveMind-Bench}, as summarized in Table \ref{tab:comparison_WaveMind_bench}. The assessment compares three configurations: random EEG input, real EEG input, and real EEG with RAG module integration. 
    
    The performance of real EEG outperforms random baselines across all datasets, which proves the effectiveness of instruction tuning. The RAG module improves classification accuracy for most cases, particularly in cognitive tasks and large options MCQs. For example, on the 40-class THING-EEG task, accuracy nearly doubled from 0.122 to 0.250, while ImageNet-EEG showed a modest gain from 0.574 to 0.603.

    \begin{table*}[h!]
        \begin{tabularx}{\textwidth}{p{2.5cm} *{10}{X}}
            \toprule
            \multirow{2}{*}{\textbf{Dataset}} & \multirow{2}{*}{\textbf{$k$}} & \multicolumn{3}{c}{\textbf{Random EEG}} & \multicolumn{3}{c}{\textbf{Real EEG}} & \multicolumn{3}{c}{\textbf{EEG w/RAG\textsuperscript{\dag}}} \\
            \cmidrule(lr){3-5} \cmidrule(lr){6-8} \cmidrule(lr){9-11}
             & & 2 & 4 & $k$ & 2 & 4 & $k$ & 2 & 4 & $k$ \\
            \midrule
            TUEV  & 6 & 0.434 & 0.240 & 0.159 & 0.940 & 0.867 & 0.888 & 0.925 & 0.890 & 0.904 \\
            TUAB  & 2 & 0.501 & / & / & 0.736 & / & / & 0.741 & / & / \\
            SEED  & 4 & 0.515 & / & 0.335 & 0.684 & / & 0.543 & 0.676 & / & 0.529 \\
            ImageNet-EEG  & 40 & 0.507 & 0.244 & 0.021 & 0.914 & 0.853 & 0.574 & 0.937 & 0.887 & 0.603 \\
            THING-EEG & 40 & 0.474 & 0.243 & 0.027 & 0.760 & 0.554 & 0.122 & 0.869 & 0.721 & 0.250 \\
            \bottomrule
        \end{tabularx}
    
        \begin{minipage}{\textwidth}
        {\tiny{
            \dag: RAG module runs over \underline{all datasets} and \underline{all tasks}, retrieves 420 possible names, and ensures at least 1 name is selected from a task.
        }
        }
        \end{minipage}
        \caption{\textbf{Averaged Classification Result on WaveMind-Bench.} The weight accuracy over $k$ options is reported, where each question consists of 1 correct and $k$-1 wrong options. The model is asked to output the letter represented by the correct options. }
        \label{tab:comparison_WaveMind_bench}
    \end{table*}

    \paragraph{Subject-Independent Evaluation}

    To assess the model in real-world scenarios, we perform out-of-domain evaluation on \textit{THING-EEG} shown in Table \ref{tab:zero_shot}. As expected, the model result decreases when encountering untrained \textit{sub-10}. Notably, we observe the higher accuracy on zero-shot than closed-set, which is due to the nature of the closed-set categories' imbalance in the THING-EEG dataset. In short, our foundation model maintained consistent and accurate decoding performance whether encountering unseen object categories or untrained subjects.

    \begin{table}[h!]
        \centering
        \footnotesize

        \begin{tabular}{l| c ccc ccc}
            \toprule
             & & \multicolumn{3}{c}{\textbf{Closed-set (1573 class)}} & \multicolumn{3}{c}{\textbf{Zero-shot (200 clsss)}} \\
            \cmidrule(lr){3-5} \cmidrule(lr){6-8}
            & $k$ & 2 & 4 & 40 & 2 & 4 & 40 \\ 
            \midrule
             \textbf{Group}& Chance & 0.500 & 0.250 & 0.033 & 0.500 & 0.250 & 0.033 \\
            \midrule
            \multirow{2}{*}{Real EEG} & Subject-Dependent & 0.728 & 0.504 & 0.096 & 0.756 & 0.574 & 0.128 \\ 
            & Subject-Independent & 0.680 & 0.419 & 0.074 & 0.689 & 0.442 & 0.058 \\ 

            \midrule
            
            \multirow{2}{*}{EEG w/RAG} & Subject-Dependent & 0.786 & 0.627 & 0.182 & 0.862 & 0.732 & 0.243 \\ 
            & Subject-Independent & 0.698 & 0.492 & 0.108 & 0.761 & 0.578 & 0.159 \\ 
            \bottomrule
        \end{tabular}
        \caption{\textbf{K-way Classification Performance on THING-EEG Dataset.}}
        \label{tab:zero_shot}
    \end{table}


    \subsubsection{Conversational Assessment}

    This section is designed to evaluate the model's open-ended conversational capabilities at the sentence level, specifically addressing predefined open-ended questions established by human evaluators.

\begin{figure}[htbp]
        \centering
        \includegraphics[width=0.9\textwidth]{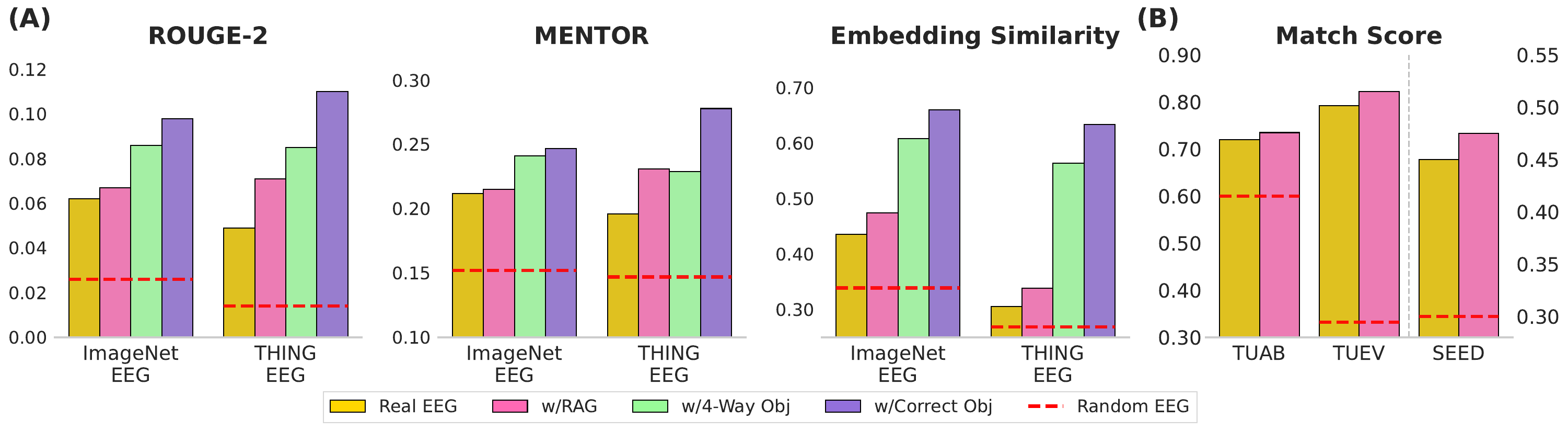}
        \caption{\textbf{Conversational Assessment with Varied Cue Granularity} (A) Cognitive evaluation where WaveMind's responses are compared with image captions from visual stimuli. The "w/Obj" indicates that the model is provided with an object cue consisting of k possible options, including one correct answer, where this object cue is directly incorporated into the input prompts; (B) Brain State evaluation, where GPT-4o is adopted to determine whether WaveMind's responses contain the correct category from clinical annotation.}
        \label{fig:v-task}
    \end{figure}

    \paragraph{Brain Cognition Understanding (EEG-Image)}
    We first evaluate \textit{Wavemind} to identify the visual content participants are currently viewing, using image captions from the visual stimuli as ground truth. To investigate how prompt granularity affects model generation, we tested three distinct levels of granularity cues, including RAG. As shown in Figure \ref{fig:v-task}, the result without any cues outperforms the random baseline, proving that the synthetic instruction effectively enhances the model's comprehension ability. Crucially, performance improves progressively as the granularity increases.

    \paragraph{Brain State Understanding (EEG-Text)}
    With the lack of sentence-level ground truth but category annotations paired with EEG, we utilize GPT-4o to determine if WaveMind's responses imply or contain phase-based annotations in binary judgments. Figure \ref{fig:v-task} demonstrates that WaveMind successfully understood the information carried by the Brain State brain activity in EEG and explained it in the form of natural language, confirming the critical role of genuine EEG input in generating relevant outputs. Meanwhile, RAG has been proven to improve the quality of generation.

\section{In-Depth Analysis}
    
    Although WaveMind has demonstrated outstanding performance, we still need to verify the superiority of its module designs through systematic experiments. Firstly, we will conduct a comprehensive performance comparison analysis between the EEG Encoder and existing mainstream methods (Section~\ref{sec:eeg-perception-on-encoder}). Secondly, through rigorous ablation experiments, we verify the optimality of each component of the model in stages (Section~\ref{sec:ablation}). Finally, we select typical dialogue cases for in-depth analysis to visually demonstrate the system's human-computer interaction capabilities (Section~\ref{sec:case-study}).

    \subsection{Encoder Perception}
    \label{sec:eeg-perception-on-encoder}
        We perform feature retrieval evaluation in Table \ref{tab:stage_I performance}. On the THING-EEG dataset, it achieves significantly better results compared to baseline models. Meanwhile, it consistently demonstrates top-3 performance on brain-state datasets, such as SEED and TUEV. Our ATMM shows remarkable competitiveness compared to alternatives, emerging as the most appropriate foundation for EEG-MLLM construction.

    
    \begin{table*}[h]
        \centering
        \small
        \renewcommand{\arraystretch}{0.95}
        \renewcommand{\tabularxcolumn}[1]{>{\centering\arraybackslash}m{#1}} 
        
        \begin{tabularx}{\textwidth}{l *{9}{X}} 
        \toprule
        
        Dataset & \multicolumn{3}{c}{\textbf{THING-EEG}} & \multicolumn{3}{c}{\textbf{ImageNet-EEG}} & \multicolumn{1}{c}{\textbf{SEED}} & \multicolumn{1}{c}{\textbf{TUAB}} & \multicolumn{1}{c}{\textbf{TUEV}} \\
        
        Category & \multicolumn{3}{c}{200} & \multicolumn{3}{c}{40} & \multicolumn{1}{c}{3} & \multicolumn{1}{c}{2} & \multicolumn{1}{c}{6} \\
        
        \midrule
        
        Validation Method & \multicolumn{3}{c}{ZS} & \multicolumn{3}{c}{CS} & CS & CS & CS \\
        
        \cmidrule(lr){2-4} \cmidrule(lr){5-7} \cmidrule(lr){8-10} 
        $k$  & 2 & 4 & 10 & 2 & 4 & 10 & 3 & 2 & 6 \\
        \midrule
        Chance          & 0.500 & 0.250 & 0.100 & 0.500 & 0.250 & 0.100 & 0.333 & 0.500 & 0.166 \\
        \midrule
        EEGITNet        & 0.716 & 0.478& 0.251& \textit{0.949} & \textit{0.872} & \textit{0.731} & 0.395 & \underline{0.722} & 0.724 \\
        MLP             & 0.780 & 0.582 & 0.361 & 0.554 & 0.293 & 0.138 & 0.354 & 0.692 & 0.687 \\
        ShallowFBCSPNet  & \textit{0.796} &\textit{0.575} & \textit{0.364} & 0.826 & 0.645 & 0.446 & 0.429 & 0.688 & 0.804 \\
        NICE          & 0.753 & 0.538 & 0.299 & 0.946 & 0.876 & 0.776 & 0.421 & 0.700 & 0.782 \\
        ATMS          & \underline{0.878} & \underline{0.767} & \underline{0.556} & 0.933 & 0.845 & 0.737 & 0.442 & 0.714 & \underline{0.893} \\
        \midrule
        \rowcolor{lightblue} \textbf{Ours (ATMM)}   & \textbf{0.909} & \textbf{0.786} & \textbf{0.631} & \underline{0.951} & \underline{0.881} & \underline{0.799} & \underline{0.493} & \textbf{0.720} & \textit{0.880} \\
        \bottomrule
        \end{tabularx}
        
        \caption{\textbf{EEG Perception Result on Encoder.} Feature retrieval accuracy is reported. Given an EEG segment, the encoder is asked to select the most similar feature from $k$ features, which consist of 1 correct and the remaining wrong. CS denotes closed-set validation, and ZS denotes zero-shot validation. SD denotes subject-dependent, SI denotes subject-independent. \textbf{BOLD} denotes Top-1, \underline{underline} denotes Top-2 and \textit{italic} denotes Top-3.}
        
        \label{tab:stage_I performance}
    \end{table*}

    \subsection{Ablation Study}
    \label{sec:ablation}
    

     \subsubsection{Encoder Alignment Matters (Stage I)}
    
            As shown in Table \ref{tab: training_rep}, when WaveMind uses a randomly initialized encoder for SFT training, its MCQ result is not significantly higher than the random guess level, far from achieving effective classification performance. This result confirms that the proposed method is crucial for constructing effective unified representations and is a necessary foundation for the model to achieve high performance.

    \subsubsection{Cold-Start Matters (Stage II)}
    Compared to directly training without a cold start, classification results gains are under different encoder settings and datasets.
    This indicates that the cold start enables the model to initially grasp the recognition ability in the CLIP space before EEG domain fine-tuning, which helps alleviate optimization bias in the early stages of the model and promotes faster convergence to better solutions under different training paradigms.

    \subsubsection{Encoder Should Freeze (Stage III)}
    Table \ref{tab: training_rep} shows that the unfrozen encoder demonstrates consistent results compared to the frozen encoder. Given the advantages of using RAG with the frozen encoder, the frozen encoder is the better choice.

        \begin{table}[h]
            \centering
            \scriptsize
            \resizebox{0.8\textwidth}{!}{
            \begin{tabular}{c c c |cccccc} 
            \toprule
             \multicolumn{2}{l}{\textbf{Encoder Alignment}} & \multirow{2}{*}{\makecell{\textbf{Cold}\\\textbf{Start}}}  & \multirow{2}{*}{\textbf{TUEV}} & \multirow{2}{*}{\textbf{SEED}} & \multirow{2}{*}{\textbf{TUAB}} & \multirow{2}{*}{\makecell{\textbf{ImageNet}\\\textbf{EEG}}} & \multirow{2}{*}{\makecell{\textbf{THING}\\\textbf{EEG}}} \\
             \cmidrule{1-2}
             Random & Freeze &  &  &  &  & &  \\
            \midrule
            \multicolumn{3}{l|}{Chance} & 0.166 & 0.333 & 0.500 & 0.020 & 0.020 \\
            
            \midrule
            
            \ding{51} & \ding{55} & \ding{51}& 0.200 & 0.296 & 0.449 & 0.026 & 0.020\\  
            \ding{55} & \ding{55} & \ding{51}& \underline{0.860} & \textbf{0.515} & \textbf{0.742} & \textbf{0.555} & \underline{0.103}\\

            \ding{55} & \ding{55} & \ding{55}& 0.641 & 0.494 & 0.686 & 0.025 & 0.030\\ 
            \ding{55} & \ding{51} & \ding{55}& 0.856 & 0.476 & 0.720 & 0.363 & 0.092\\

    
            \rowcolor{lightblue} \ding{55} & \ding{51} & \ding{51}& \textbf{0.873} & \underline{0.498} & \textbf{0.737} & \underline{0.390} &\textbf{0.108}\\
            \bottomrule
            \end{tabular}
            }
            \caption{\textbf{Ablation Study over Training Recipe.} The result in \textit{n} options is reported, where \textit{n} is the class of datasets.}
            \label{tab: training_rep}
        \end{table}

    \subsubsection{Scaling Helps Diversity and Ability (Stage III)}
    Our experiments in Table \ref{tab:linguistic_diversity} with varying training data sizes demonstrate that the full data consistently outperforms partial data across linguistic diversity and classification ability. This proves that Deduplication and quality control are necessary, which enables MLLM to learn more evenly distributed language patterns, significantly improving the richness of generated content.


    \begin{table}[h]
        \centering
        \small
        \begin{tabular}{l c c c c c |c c c}
            \toprule
            \multirow{3}{*}{\textbf{Data}}& \multicolumn{5}{c}{\textbf{Linguistic Diversity}} & \multicolumn{3}{c}{\textbf{Classification Ability}}\\
            \cmidrule(lr){2-6}  \cmidrule(lr){7-9}
            \multirow{2}{*}{\textbf{}} & \multicolumn{2}{c}{\textbf{Distinct $\uparrow$}} & \multirow{2}{*}{\makecell{\textbf{Jaccard}\\\textbf{Diversity}}$\uparrow$} & \multirow{2}{*}{\makecell{\textbf{Cosine}\\\textbf{Diversity}}$\uparrow$} & \multirow{2}{*}{\makecell{\textbf{Self}\\\textbf{BLEU}}$\downarrow$} 
            &  \multirow{2}{*}{\makecell{\textbf{ImageNet}\\\textbf{EEG}}$\uparrow$} 
            &  \multirow{2}{*}{\makecell{\textbf{THING}\\\textbf{EEG}}$\uparrow$} 
            &  \multirow{2}{*}{\makecell{\textbf{TUAB}}$\uparrow$} 
            
            \\
            \cmidrule{2-3}
            & 1-gram & 2-gram & & & \\
            \midrule
            \textbf{5\%}   & 0.100 & 0.317 & 0.857 & 0.884 & 0.840 &0.694& 0.574&0.528\\
            \textbf{10\%}  & 0.094 & 0.306 & \textbf{0.865} & 0.896 & 0.838 &0.762&0.632&0.628\\
            \rowcolor{lightblue}\textbf{Full}  & \textbf{0.117} & \textbf{0.353} & 0.862 & \textbf{0.903} & \textbf{0.799} &\textbf{0.936}& \textbf{0.694}& \textbf{0.722}\\
            \bottomrule
        \end{tabular}
        \caption{\textbf{Ablation on training using different scales of data.} 2-way accuracy of each dataset is reported in the classification ability. }
        \label{tab:linguistic_diversity}
    \end{table}

     \subsection{Case Study}
     \label{sec:case-study}
    
\begin{figure}[htbp]
            \centering
            \includegraphics[width=0.7\linewidth]{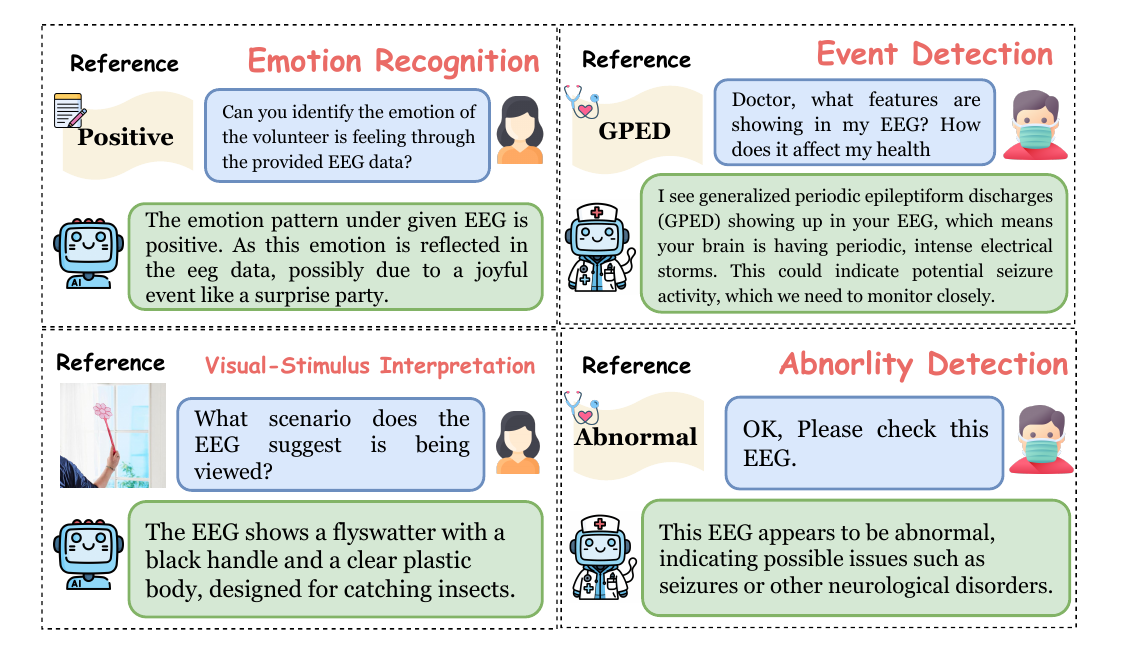}
            \caption{\textbf{Case Study of Real Conversation between Human and WaveMind.}}
            \label{fig:overall_demo}
        \end{figure}

        Figure \ref{fig:overall_demo} shows a real conversation between a human and WaveMind, demonstrating the model's strong understanding of common situations. Importantly, WaveMind performs well even with open-ended prompts lacking specific instructions, reflecting its ability to infer user intent. These examples indicate that WaveMind can effectively identify and categorize tasks, validating both our synthetic instruction approach and the success of instruction tuning. More cases and scenarios are demonstrated in Appendix \ref{sec:Open Chat Demostration}.

\section{Conclusion}


In this study, we propose \textit{WaveMind}, a novel alignment framework for EEG-MLLM systems. In general, \textit{WaveMind} bridges the paradigm of EEG foundational models (enabling cross-task generalization) and EEG conversational models (facilitating EEG interpretation). Leveraging a unified representation space, our framework enhances downstream tasks through three perspectives: (1) improved conversational ability, (2) robust EEG signal awareness, and (3) utilization of upstream training data. Notably, existing approaches such as NeuroLM \cite{jiang2025neurolm} and Thought2EEG \cite{mishraThought2TextTextGeneration2024} lack these integrated capabilities. Comprehensive comparisons with related work are provided in Appendix~\ref{sec:related_work}.

\paragraph{Technical Insights} Training MLLMs within this unified latent space unlocks two key advantages: (1) it enables cross-activity and cross-task EEG perception (Tab. \ref{tab:stage_I performance}); (2) it supports cold-start tuning by leveraging out-of-domain data, dramatically improving model adaptability before EEG-domain tuning (Tab. \ref{tab: training_rep}). Notably, our experiments show that synthesized instruction data effectively develops the model's ability to identify (Tab. \ref{tab:comparison_WaveMind_bench}) and interpret (Tab. \ref{fig:v-task}) EEG. Additionally, rigorous data engineering (Fig. \ref{fig:trainingStage}) practices play a crucial role in enhancing both data diversity and model performance (Tab. \ref{tab:Multi-modilities helps}).

\paragraph{Neuroscience Insights} Our analysis reveals complementary neural patterns between cognitive and Brain State activities (Tabs.\ref{tab:pivot_study} \ref{tab:Multi-modilities helps}). This complementarity underlies the brain's ability to maintain cognitive clarity in dynamic environments while remaining sensitive to internal physiological states. Our result demonstrated specific manifestation patterns of this complementarity in the EEG. These findings provide critical neural-level insights for deepening our understanding of integrated brain function.


\clearpage
\bibliographystyle{unsrt}
\bibliography{reference}


\clearpage
\appendix
\begin{center}
\Large\textbf{Supplementary Material}
\end{center}

\section{More Results}
\subsection{Ablation on Selected LLM Backbones and Encoders}

\begin{table}[h!]
        \centering
        \begin{tabularx}{0.8\textwidth}{l|XXXcc} 
        \toprule
        & \textbf{TUEV} & \textbf{SEED} & \textbf{TUAB} & \makecell{\textbf{ImageNet}\\\textbf{EEG}} & \makecell{\textbf{THING}\\\textbf{EEG}} \\
        \midrule
        
        \multicolumn{1}{c }{Chance} & 0.166 & 0.333 & 0.500 & 0.020 & 0.020 \\

        \midrule
        \multicolumn{1}{l|}{\textbf{\textit{LLM Backbone\textsuperscript{\dag}}}}\\
        
        \multicolumn{1}{c|}{Qwen2.5-Insturct 7B} & 0.758 & 0.542 & 0.735 & 0.478 &\textbf{ 0.131} \\  
        \multicolumn{1}{c|}{Mistral-Instruct-0.3 7B} & \textbf{0.937} & 0.440 & 0.707 & 0.454 &\textbf{ 0.131} \\
        \rowcolor{lightblue}\multicolumn{1}{c|}{\textit{Vicuna-1.5 7B}} &0.860 & \textbf{0.515} & \textbf{0.742} & \textbf{0.555} & 0.103  \\ 
        \midrule
        \multicolumn{1}{l|}{\textbf{\textit{EEG Encoder\textsuperscript{\ddag}}}}\\
        \multicolumn{1}{c|}{ChannelNet} & 0.390 & 0.411 & 0.634 & 0.049 & 0.050 \\  
        \rowcolor{lightblue}\multicolumn{1}{c|}{ATMM} & \textbf{0.860} & \textbf{0.515} & \textbf{0.742} & \textbf{0.555} & \textbf{0.103} \\
        
        \bottomrule
        \end{tabularx}

         \begin{minipage}{0.8\textwidth}
        {\renewcommand{\baselinestretch}{0.1}
        \tiny
            \textsuperscript{\dag} The train is under the \textit{ATMM} encoder with different backbone.\textsuperscript{\ddag} The train is under the \textit{Vicuna} backbone with different encoder.
        }
        \end{minipage}
        
        \caption{\textbf{Ablation Study over Different Backbone and Encoder}}
        \label{tab: training_backbone}
        
    \end{table}

Table \ref{tab: training_backbone} indicates that the WaveMind model using different backbone architectures did not show significant differences in performance metrics. This result indicates that the key bottleneck currently constraining model performance is not the Large Language Model (LLM) backbone network itself, but rather may stem from non-model architecture factors such as the degree of data engineering optimization or data quality characteristics. 

As for encoder ablation, ATMM, as the Encoder, achieved better results, especially significantly outperforming other encoders in the ImageNet and THING-EEG datasets.
Therefore, our proposed ATMM is the best choice for WaveMind.

\subsection{Ablation on Selected Loss Function in Representation Alignment}

\label{sec:loss_vs}

\begin{table}[h!]
\renewcommand{\arraystretch}{1.1}
\centering

\begin{tabularx}{0.8\textwidth}{lXXcc}
\toprule
 & \textbf{Class} & \textbf{K-way} & \textbf{CE+InfoNCE} & \textbf{InfoNCE} \\ \midrule
\textit{\textbf{Brain State (Text-EEG)}}\\
TUEV & 6 & 6 & \textbf{0.900} & 0.880 \\ 
TUAB & 2 & 2 & \textbf{0.719} & 0.707 \\ 
SEED & 3 & 3 & 0.462 & \textbf{0.492} \\ 
\midrule
\textit{\textbf{Brain Cognition(Image-EEG)}}\\
ImageNet-EEG & 40 & 40 & 0.617 & \textbf{0.829} \\ 
\cmidrule{2-5}
\multirow{4}{*}{THNING-EEG} & \multirow{4}{*}{200} & 2 & 0.902 & \textbf{0.909} \\ 

 & & 4 & 0.783 & \textbf{0.789} \\ 
 & & 10 & 0.578 & \textbf{0.620} \\ 
 & & 200 & 0.125 & \textbf{0.146} \\ \bottomrule
 
\end{tabularx}
\caption{\textbf{Ablation on Selected Loss Function.} K-way Weight Feature Retrieval Accuracy is reported, where K represents the number of candidate features provided to the encoder}
\label{tab:loss_vs}
\end{table}

\subsection{Supervision Modalities are Complementary}
    As shown in Table \ref{tab:Multi-modilities helps}, we perform the SFT from WaveMind-Instruct. The result shows improved performance across all downstream tasks after co-training of cognitive and Brain State tasks. The co-training enhances the model's representation of complex EEG signals and enables complementary knowledge transfer via multimodal interactions.

    \begin{table}[h!]
        \centering
        \small
        \renewcommand{\arraystretch}{1}
        \newcolumntype{C}{>{\centering\arraybackslash}X} 
        \begin{tabularx}{0.93\columnwidth}{lccCCC}
        \toprule
         & \multicolumn{2}{c}{\textbf{Brain Cognition Task\includegraphics[scale=0.03]{Image/logo/cognitive.png}}} & \multicolumn{3}{c}{\textbf{Brain State Task\reflectbox{\includegraphics[scale=0.033]{Image/logo/non_cognitive.png}}}} \\
        \cmidrule(lr){2-3}  \cmidrule(lr){4-6}
          & ImageNet-EEG & THING-EEG & TUAB & TUEV & SEED \\ 
        \midrule
        ImageNet-EEG & 0.910 & - & - & - & -\\ 
        THING-EEG & - & \textbf{0.716} & - & - & -\\ 
        TUAB & - & - & 0.716 & - & - \\ 
        TUEV & - & - & - & 0.934 & - \\ 
        SEED & - & - & - & - & 0.680\\ 
        \midrule
        \cellcolor{lightblue}\textbf{ALL} 
        & \cellcolor{lightblue}\textbf{0.936} \textcolor{orange}{\tiny(+0.026)}& \cellcolor{lightblue}0.694 \textcolor{red}{\tiny(-0.022)}& \cellcolor{lightblue}\textbf{0.722} \textcolor{orange}{\tiny(+0.006)}& \cellcolor{lightblue}\textbf{0.951} \textcolor{orange}{\tiny(0.017)}& \cellcolor{lightblue}\textbf{0.696} \textcolor{orange}{\tiny(+0.016)}\\ 
        \bottomrule
        \end{tabularx}
        \caption{\textbf{Ablation on cross data source training.} The result in 2 options is reported, which clearly suggests that combining multiple modalities results in slightly improved performance on WaveMind-bench. More importantly, this also proves the advantages of mixing Brain State and cognitive training.}
        \label{tab:Multi-modilities helps}
    \end{table}

\section{Broader Impact Analysis}

    \subsection{Positive Impact}
    
    \paragraph{}
    This research pioneers the development of WaveMind, the world's first chat-style multilingual large language model (MLLM) specifically designed for electroencephalography (EEG) signals. This breakthrough achievement realizes cross-task and cross-scenario neural signal interpretation capabilities. Through the simultaneous release of the first comprehensive benchmarking framework for EEG MLLM, we systematically evaluate the model's performance across cognitive category recognition dimensions. This work establishes a methodological foundation for developing human-machine neural interaction paradigms, marking a foundational milestone in advancing non-invasive brain-computer interface technology toward practical applications.
    
    \subsection{Negative Impact}
    
        \paragraph{Hallucination}
        Due to training data limitations containing only coarse-grained annotations, the model demonstrates pattern recognition capabilities but lacks causal reasoning mechanisms. When confronted with attribution analysis tasks, it may generate content conflicting with established neuroscience principles. We strongly recommend restricting its application scope and prohibiting direct deployment in high-risk domains like clinical diagnostics.
        
        \paragraph{Malicious input}
        Key threats include: (1) Null data injection causing abnormal interpretations; (2) Malicious instructions inducing response biases. Our dual-layer protection mechanism includes: (a) Embedding 1\% non-EEG instruction in training data to develop rejection capabilities; (b) Deploying a pre-detection module to intercept attribution-related queries.
        
        \paragraph{Out-domain Generalization}
        While WaveMind demonstrates zero-shot and cross-subject capabilities on THING-EEG, its generalization to other tasks and datasets cannot be guaranteed. Inappropriate EEG channel configurations and sampling rates may cause performance degradation. To address this, we developed a preprocessor module to standardize input formats.
        
        \paragraph{Bias}
        Potential biases may originate from: (1) EEG encoder (ATMM); (2) Backbones (LLaMA/Vicuna/ Qwen); (3) Training data (Llava\_Instruct and WaveMind-Instruct). Despite implementing diverse quality control measures, complete bias elimination cannot be guaranteed, potentially leading to biased outputs or unfair representations of diverse content. The bias of the model originates from the structure and respective training processes themselve. For training data, we have taken proactive measures to prevent factual errors, see Appendix \ref{sec:human_eval}.
        
        \paragraph{Energy}
        While current energy consumption is not critical, expanding the pretraining dataset or increasing model scale could become significant concerns requiring optimization strategies.
        
        \paragraph{Evaluation}
        Current benchmarks are constrained by coarse annotation granularity, covering only MCQ formats. Quantitative evaluation metrics for neuroscience interpretability remain underdeveloped, necessitating future development of comprehensive assessment systems incorporating causal reasoning and cross-modal alignment dimensions.
        
    Despite these identified risks, we maintain that WaveMind's positive contributions outweigh its limitations, as it provides valuable methodological support for the research community.

\section{Open Chat Demostration}
\label{sec:Open Chat Demostration}
        \begin{figure}[h!]
            \centering
            \includegraphics[width=0.8\linewidth]{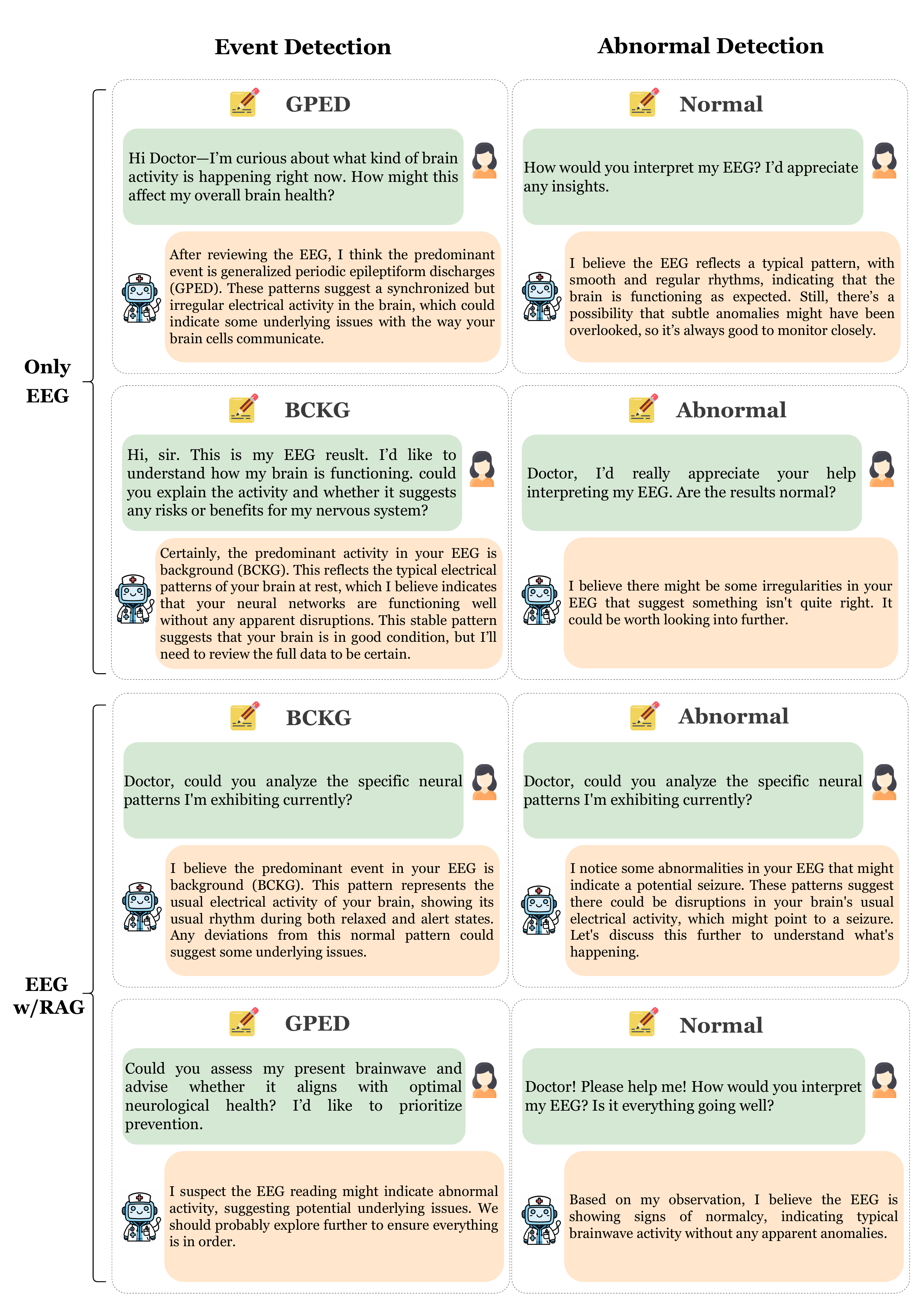}
            \caption{\textbf{Interpretation on Brain State Activity over \textit{Consultation} Scenario}}
            \label{fig:enter-label}
        \end{figure}

        \begin{figure}[h!]
            \centering
            \includegraphics[width=1\linewidth]{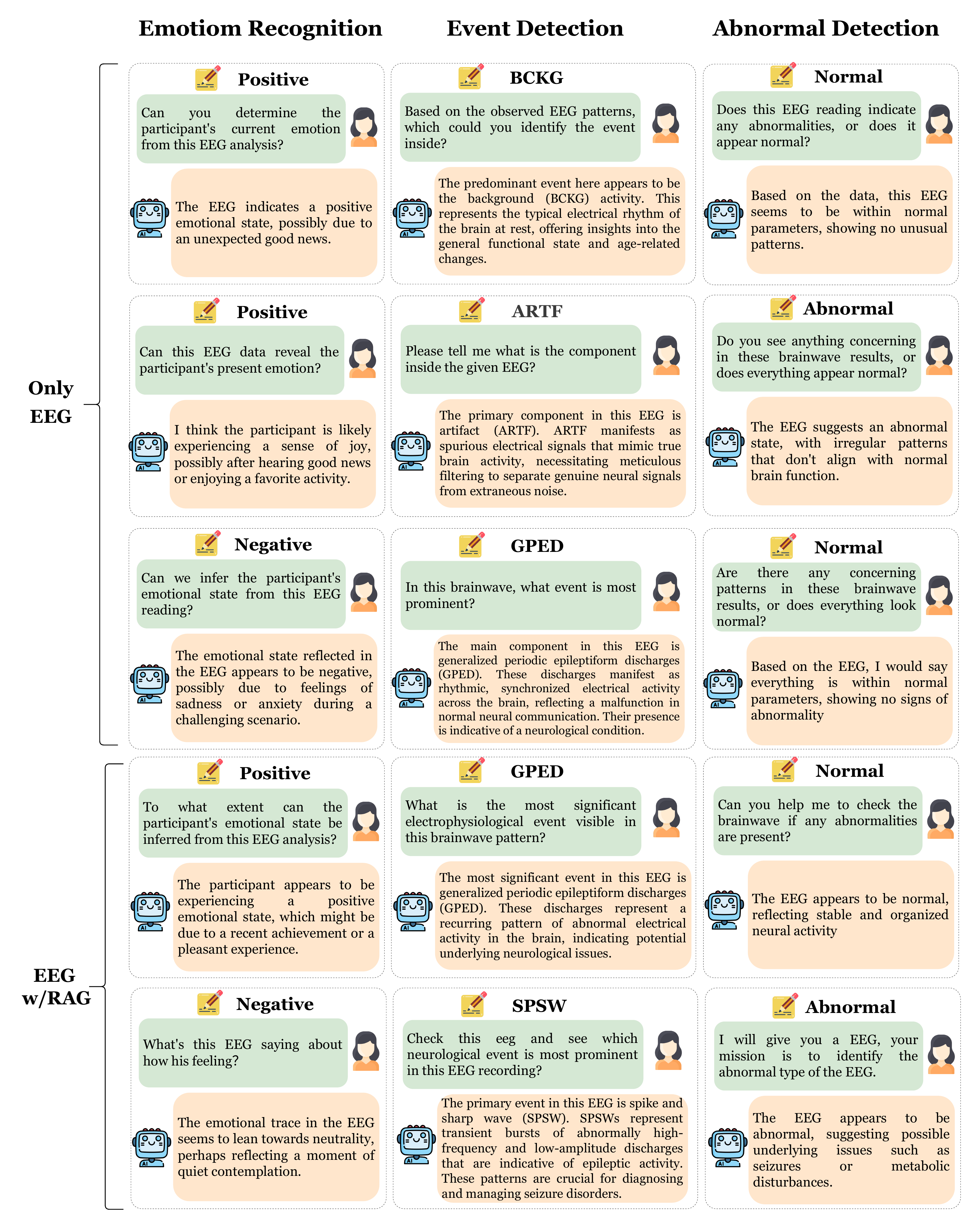}
            \caption{\textbf{Interpretation on Brain State Activity over \textit{Analysis} Scenario}}
            \label{fig:demo_ei_appendix}
        \end{figure}

        \begin{table}[h!]
        \small
        \centering
        \renewcommand{\tabularxcolumn}[1]{>{\centering\arraybackslash}m{#1}}
        \begin{tabularx}{\textwidth}{>{\hsize=0.25\hsize}X>{\hsize=0.75\hsize}X}
        \toprule
        \textbf{Label} & \textbf{Content} \\
        \midrule
         \includegraphics[width=0.25\linewidth]{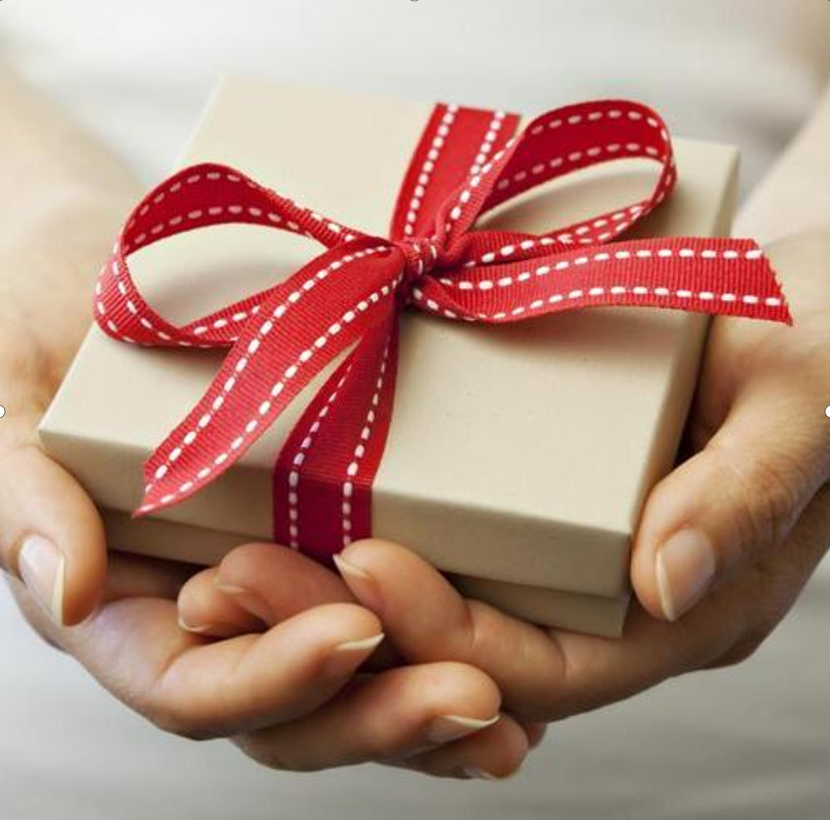} & 
        \begin{itemize}
            \item \textbf{Input} EEG w/Correct Label
            \item \textbf{Object} Pizza
            \item \textbf{Ground Truth} A pair of hands gently holds a small, rectangular gift wrapped in brown paper and tied with a red ribbon adorned with white dots.
            \item \textbf{LLM Response} \textcolor{darkgreen}{A gift box with a festive design, ready for giving. }
        \end{itemize}
         \\
         
        \midrule
         \includegraphics[width=0.25\linewidth]{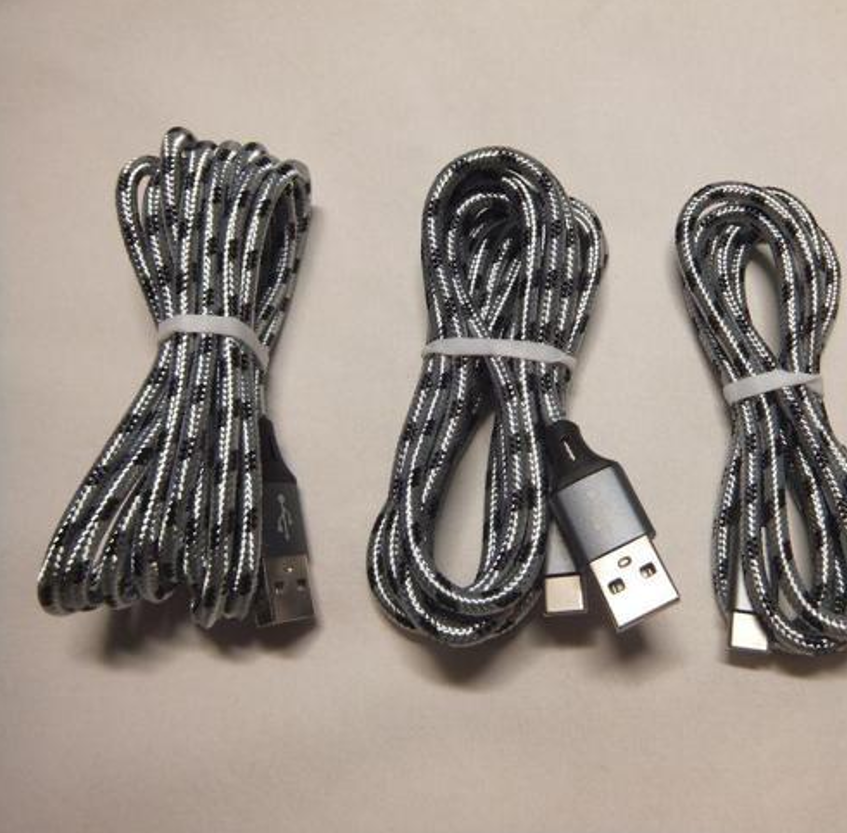} & 
        \begin{itemize}
            \item \textbf{Input} EEG w/4 way hint
            \item \textbf{Object} Cable
            \item \textbf{Ground Truth} three coiled USB cables, each with a metallic braided design and a white connector at one end. The cables are neatly bundled together and placed side by side against a plain background.
            \item \textbf{LLM Response} \textcolor{darkgreen}{a cable with a metallic sheath and a black outer layer, resting on a wooden surface.}
        \end{itemize}
         \\

        \midrule
        \includegraphics[width=0.25\linewidth]{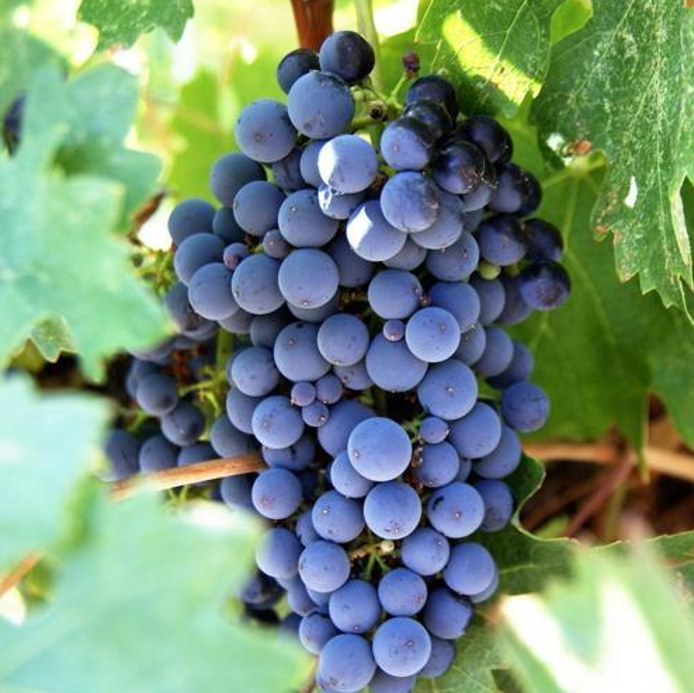} & 
        \begin{itemize}
            \item \textbf{Input} EEG w/4 way hint
            \item \textbf{Object} Grape
            \item \textbf{Ground Truth} a cluster of dark purple grapes hanging from a vine, surrounded by green leaves. The grapes appear ripe and plump, ready for harvest.
            \item \textbf{LLM Response} \textcolor{darkgreen}{a grape, with its rich purple hue and textured skin visible.}
        \end{itemize}
         \\

        \midrule
        \includegraphics[width=0.25\linewidth]{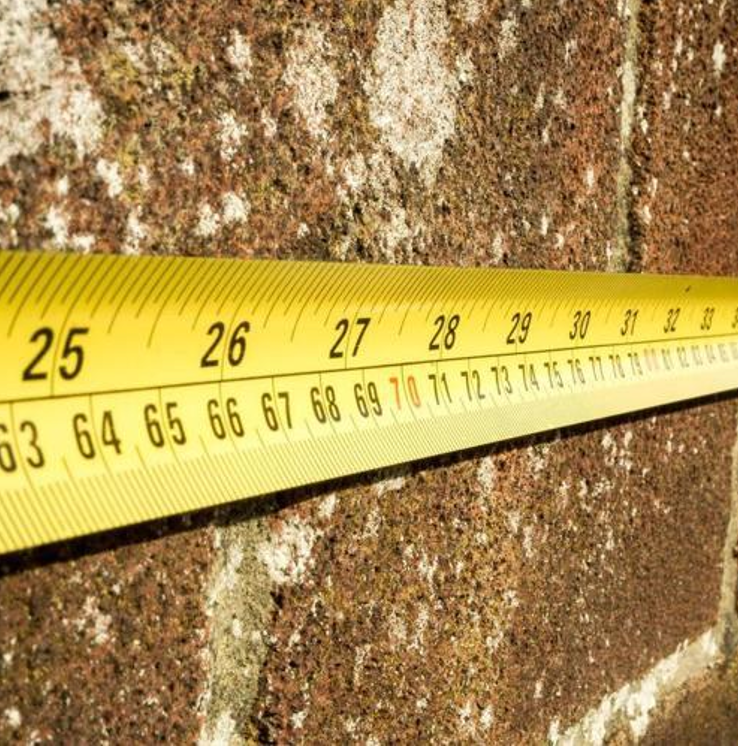} & 
        \begin{itemize}
            \item \textbf{Input} EEG w/4 way hint
            \item \textbf{Object} Ruler
            \item \textbf{Ground Truth} A yellow tape measure lies against a textured brick wall, displaying measurements from 25 to 33 inches. The numbers are marked, with some sections highlighted in red for emphasis.
            \item \textbf{LLM Response} \textcolor{red}{a beaver with its tail flipped up, revealing its muscular legs and powerful hindquarters.}
        \end{itemize}
         \\

        \midrule
        \includegraphics[width=0.25\linewidth]{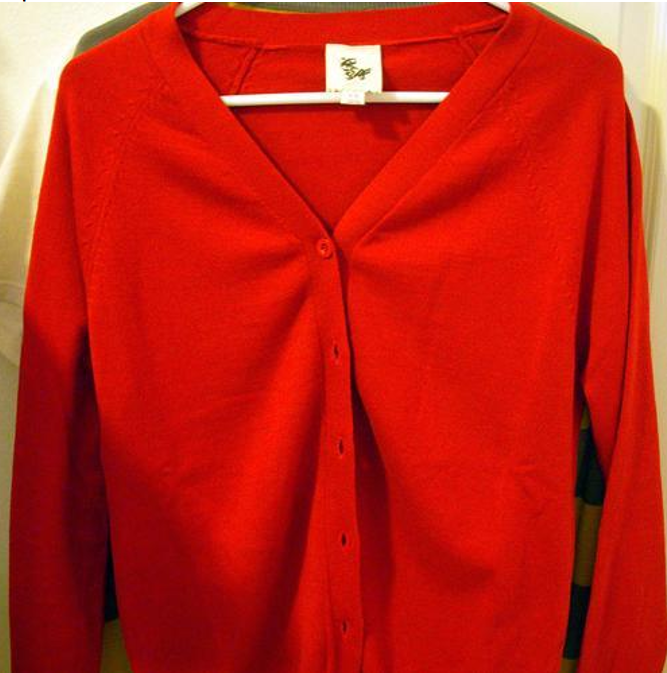} & 
        \begin{itemize}
            \item \textbf{Input} EEG w/4 way hint
            \item \textbf{Object} Cardigan
            \item \textbf{Ground Truth} a red cardigan hanging on a hanger, featuring a V-neck design and a row of buttons down the front. The cardigan appears to be made of a soft, knit material.
            \item \textbf{LLM Response} \textcolor{red}{a person wearing a \textcolor{darkgreen}{red} leotard with a black bow. The leotard has a sleek, fitted design with a high neckline and a short, flowing skirt. The bow is tied neatly at the back, adding a touch of elegance to the outfit.}
        \end{itemize}
         \\

        \midrule
        \includegraphics[width=0.25\linewidth]{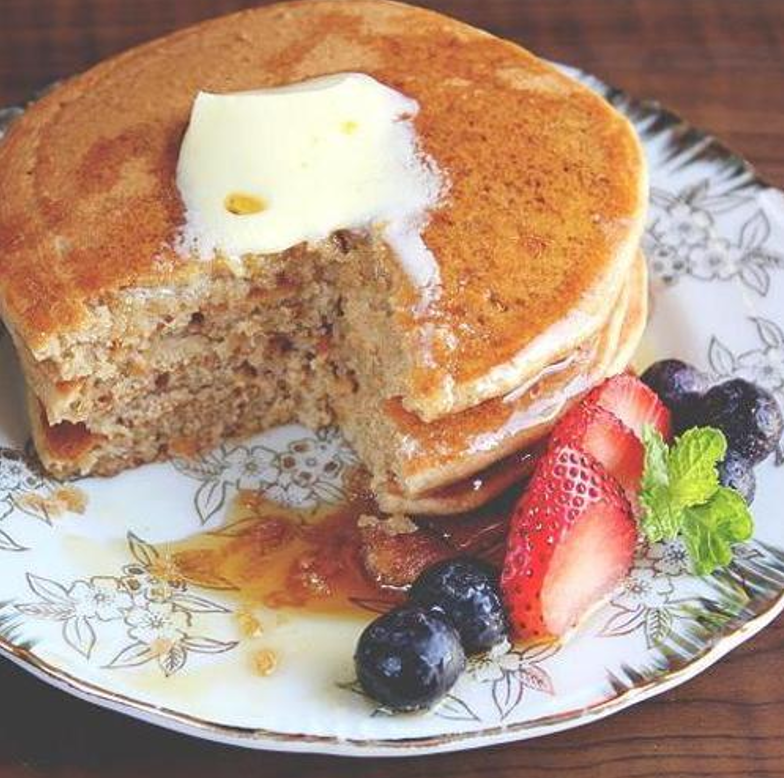} & 
        \begin{itemize}
            \item \textbf{Input} EEG w/4 way hint
            \item \textbf{Object} Cake
            \item \textbf{Ground Truth} A stack of pancakes is served on a decorative plate, topped with a pat of butter and drizzled with syrup. The plate also features fresh strawberries and blueberries as garnish.
            \item \textbf{LLM Response} \textcolor{darkgreen}{a delicious-looking burrito with a golden-brown crust, filled with a variety of ingredients including beans, cheese, and meat.}
        \end{itemize}
         \\

        \bottomrule
        \end{tabularx}
        \caption{\textbf{Interpretation on Brain Cognition over \textit{THING-EEG} dataset}}
        \label{tab:example}
        \end{table}

        \begin{table}[h!]
        \centering
        \small
        \renewcommand{\tabularxcolumn}[1]{>{\centering\arraybackslash}m{#1}}
        \begin{tabularx}{\textwidth}{>{\hsize=0.25\hsize}X>{\hsize=0.75\hsize}X}
        \toprule
        \textbf{Label} & \textbf{Content} \\
        \midrule
         \includegraphics[width=0.25\linewidth]{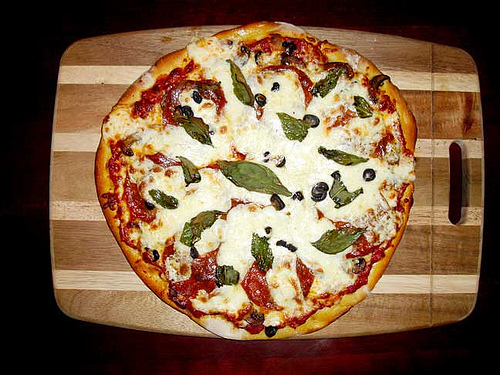} & 
        \begin{itemize}
            \item \textbf{Input} EEG w/Correct Label
            \item \textbf{Object} Pizza
            \item \textbf{Ground Truth} a freshly baked pizza topped with melted cheese, tomato sauce, black olives, and fresh basil leaves, placed on a wooden cutting board.
            \item \textbf{LLM Response} \textcolor{darkgreen}{a pizza with a golden-brown crust and generous toppings, including pepperoni and mushrooms. }
        \end{itemize}
         \\

        \midrule
         \includegraphics[width=0.25\linewidth]{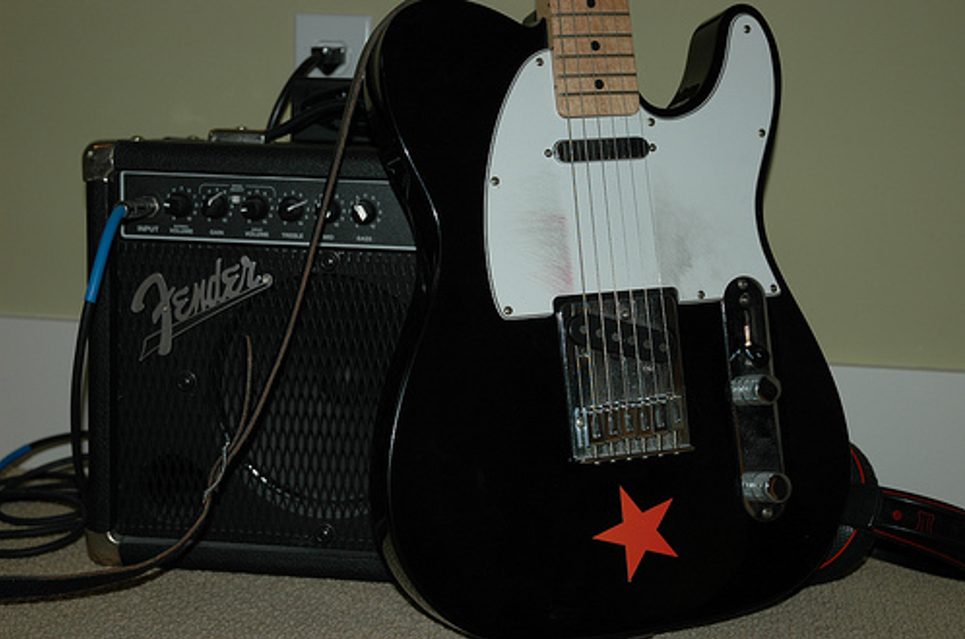} & 
        \begin{itemize}
            \item \textbf{Input} EEG w/4 way hint
            \item \textbf{Object} Guitar
            \item \textbf{Ground Truth} a black electric guitar with a red star on its body, positioned next to a Fender amplifier. The guitar has a white pickguard and a maple neck.
            \item \textbf{LLM Response} \textcolor{darkgreen}{a guitar with a dark body and a light-colored neck, positioned on a \textcolor{red}{wooden} surface.}
        \end{itemize}
         \\

         \midrule
         \includegraphics[width=0.25\linewidth]{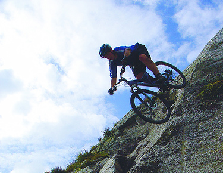} & 
        \begin{itemize}
            \item \textbf{Input} EEG w/4 way hint
            \item \textbf{Object} bike
            \item \textbf{Ground Truth} an elephant standing on a patch of grass and dirt, casting a shadow on the ground. The elephant appears to be in a natural habitat, possibly a savanna or a similar environment.
            \item \textbf{LLM Response} \textcolor{red}{a person is paddling a canoe through a river, surrounded by lush greenery and a backdrop of mountains.}
        \end{itemize}
         \\

         \midrule
         \includegraphics[width=0.25\linewidth]{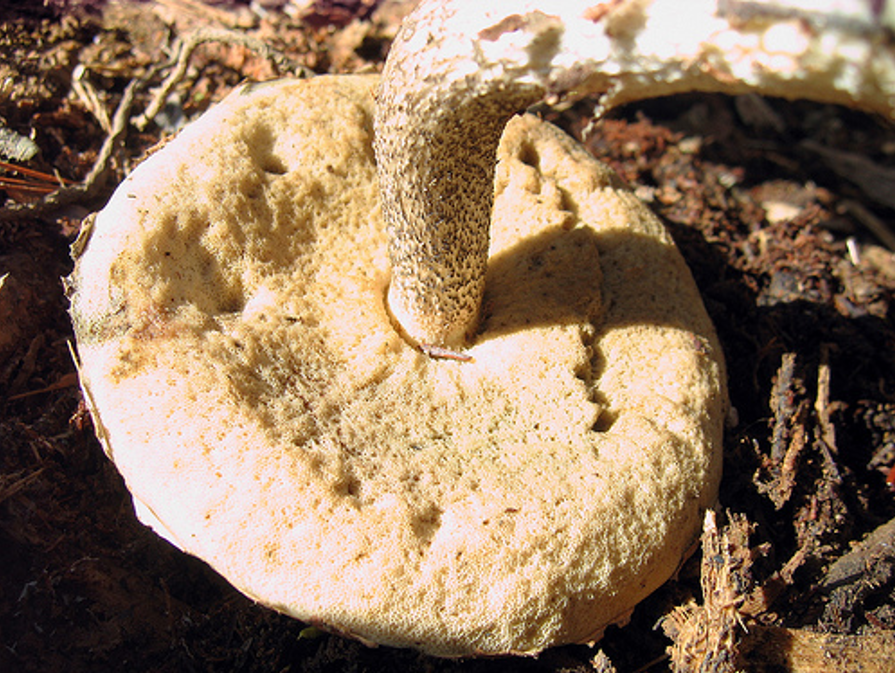} & 
        \begin{itemize}
            \item \textbf{Input} EEG w/4 way hint
            \item \textbf{Object} Mushroom
            \item \textbf{Ground Truth} a close-up of a bolete mushroom, characterized by its porous cap and central stem. The cap appears to be light-colored with a rough texture, and the stem is thick and sturdy.
            \item \textbf{LLM Response} \textcolor{darkgreen}{a close-up view of a mushroom with a dark brown cap and a white stem. The mushroom has a slightly curved cap and a stem that tapers towards the bottom.}

        \end{itemize}
         \\

         \midrule
         \includegraphics[width=0.25\linewidth]{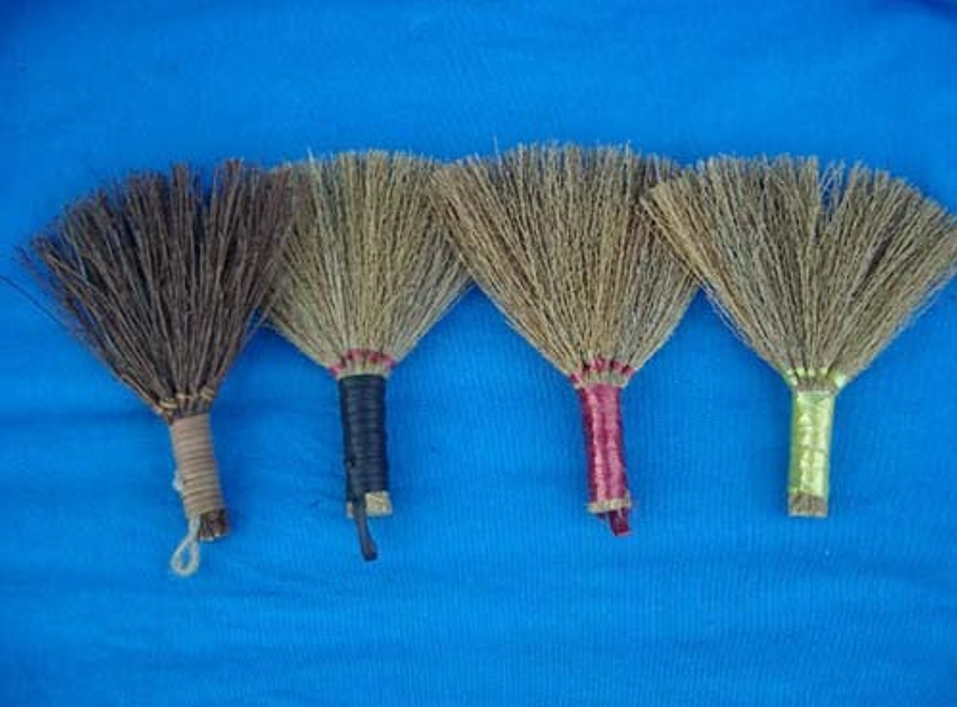} & 
        \begin{itemize}
            \item \textbf{Input} EEG w/4 way hint
            \item \textbf{Object} Four brooms with natural bristles, each featuring a different colored handle (black, brown, red, and green), arranged on a blue background.
            \item \textbf{Ground Truth} a brown leather handbag with a rectangular pocket and top handles.
            \item \textbf{LLM Response} \textcolor{red}{a close-up view of a rustic wooden spoon with a dark brown handle and a silver-colored bowl.}

        \end{itemize}
        \\

        \midrule
         \includegraphics[width=0.25\linewidth]{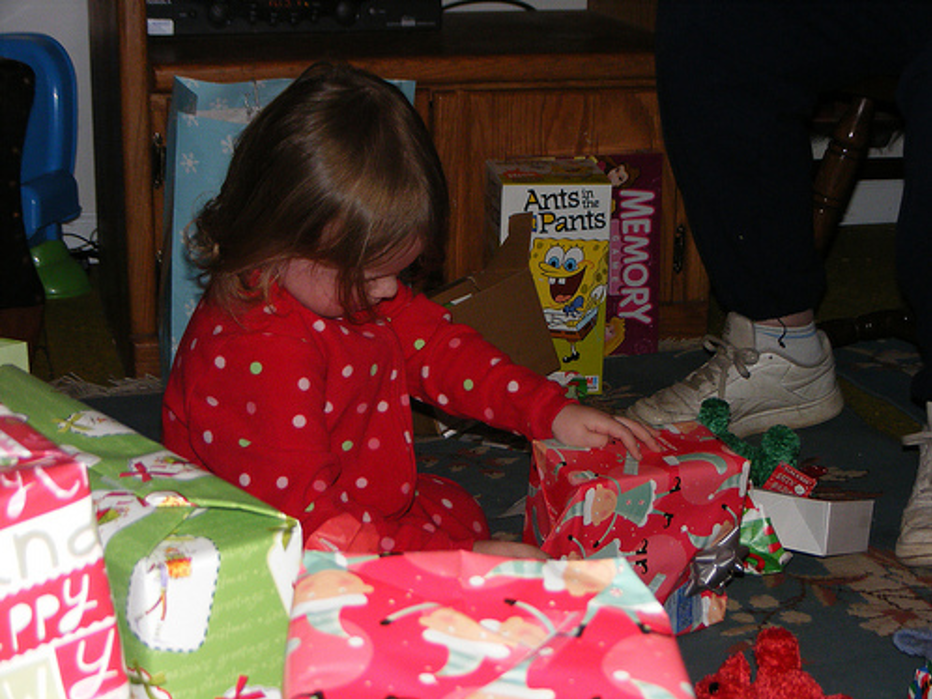} & 
        \begin{itemize}
            \item \textbf{Input} EEG w/4 way hint
            \item \textbf{Object} Child
            \item \textbf{Ground Truth} A child in a red outfit is sitting on the floor, opening wrapped presents while a cat and some holiday decorations are visible nearby.
            \item \textbf{LLM Response} \textcolor{darkgreen}{a child's face, focusing on the eyes and nose, with a neutral background.}

        \end{itemize}
         \\

        \bottomrule
        \end{tabularx}
        \caption{\textbf{Interpretation on Brain Cognition over \textit{ImageNet-EEG} dataset}}
        \label{tab:example}
        \end{table}

\clearpage

\section{Baseline EEG Encoder}
To evaluate the effectiveness of the ATMM architecture, we construct and compare baseline encoders proposed by prior work, which are as follows:

\begin{itemize}
    \item \textbf{ChannelNet} \cite{palazzoDecodingBrainRepresentations2021}: A model that processes the spatiotemporal features of the EEG signal, facilitating multimodal learning and the classification of visual representations in the brain.

    \item \textbf{NICE} \cite{songDecodingNaturalImages2024}: A self-supervised contrastive learning-based model that aligns EEG signals with natural image features to enable object recognition by decoding EEG representations of visual stimuli.
    
    
    \item \textbf{EEG-ITNet} \cite{salamiEEGITNetExplainableInception2022}: A lightweight end-to-end model based on causal dilated convolution and Inception modules for EEG signal decoding.
    
    \item \textbf{MLP} \cite{liVisualDecodingReconstruction2024}: A model solely based on MLPs and LayerNorm.
    
    \item \textbf{ShallowFBCSPNet} \cite{schirrmeisterDeepLearningConvolutional2017}: A shallow convolutional neural network that combines filter banks (FBs) and common spatial patterns (CSPs) for EEG signal classification.
    
    \item \textbf{ATM-S} \cite{liVisualDecodingReconstruction2024}: A model that integrates channel attention mechanisms, spatiotemporal convolutions, and MLP architectures to achieve visual decoding and reconstruction tasks.
\end{itemize}

Since these baselines were originally designed for specific tasks (e.g., emotion recognition, motor imagery classification, etc.), we adapted their input layers and replaced the task-specific heads to fit the data format used in this study, enabling cross-model multi-task performance comparisons.

\section{Architecture Detail}
\subsection{ATMM Encoder}
\label{ATMM Encoder}

    Our ATMM encoder is adapted from the ATM-S \cite{liRealMindZeroShotEEGBased2024}, the detail architecutre comparison is demostrated in \ref{tab:ATMSvsATMM}. To better support joint training on multiple datasets, we removed the original subject token and added multiple layers of multi-head attention (MHA). In addition, to achieve Dual-Supervision Representation Alignment, we adjusted the dimension of the output features to 768 to adapt to the CLIP space.

    Therefore, given an EEG sample $X_e$, a multi-stage processing pipeline extracts its feature vector through the following procedure. First, an enhanced Embedding Layer encodes the multi-channel EEG signals into discrete token sequences, wherein each channel undergoes independent convolution for spatial feature extraction. Subsequently, a channel attention layer employs multi-head self-attention to model cross-channel interaction relationships. Following this, the Temporal-Spatial Module jointly extracts spatiotemporal features, employing dilated convolutions to capture long-range temporal dependencies while utilizing graph convolutional networks to model spatial electrode topology. Finally, a Multilayer Perceptron (MLP) performs higher-order abstraction on the fused features, outputting a compact feature representation $\mathbf{x}' \in \mathbb{R}^{d_e}$.

    \begin{table}[htbp]
        \centering
        \begin{tabularx}{0.9\textwidth}{
            >{\bfseries}l      
            c                   
            >{\centering\arraybackslash}X 
            c                   
            >{\centering\arraybackslash}X 
            c                   
        }
            \toprule
           \multirow{2}{*}{Layer} & \multicolumn{2}{c}{\textbf{ATM-S}} & \multicolumn{2}{c}{\textbf{ATMM}} & \multirow{2}{*}{\textbf{Input Shape}} \\
            \cmidrule(r){2-3} \cmidrule(r){4-5}
            & Block & \#Parameter & Block & \#Parameter & \\
            \midrule
            Embedding & 1x & 270,848 & 1x & 270,848 & (B, C, T) \\
            \quad \normalfont{Subject Token} & \ding{51} & - & \ding{55} & - & - \\
            \quad \normalfont{Share Token} & \ding{51} & - & \ding{51} & - & - \\
            MHA & 2x & 2,631,168 & 12x & 33,615,872 & (B, C, D) \\
            TS Conv & 1x & 52,440 & 1x & 52,440 & (B, C, D) \\
            TS Agg & 1x & - & 1x & - & (B, H1, H2) \\
            MLP projector & 1x & 3,605,504 & 1x & 2,704,128 & (B, H1*H2) \\
            \midrule
            \textbf{Output Shape} & \multicolumn{2}{c}{(B, 1024)} & \multicolumn{2}{c}{(B, 768)} &  \\
            \textbf{Total} & \multicolumn{2}{c}{8,663,872}  & \multicolumn{2}{c}{37,827,648}  \\
            \bottomrule
        \end{tabularx}
        \caption{\textbf{The comparison of architecture between ATM-S and ATMM encoder}, where B, C, T represent batch size, channels, and time.}
        \label{tab:ATMSvsATMM}
    \end{table}

\subsection{RAG Module}
    \label{sec:RAG_module}
    \subsubsection{RAG Detail}
    RAG is used to enhance the WaveMind Generation. Pre-defined Supervision features and associated categories (name) from all datasets are precomputed and stored. Given EEG\_sample matches top-k features and their corresponding names based on cosine similarity, and is inserted as a piece of text into the prompt. To ensure robustness across the tasks, we enforce that at least one name is selected from each task. These names are then encoded through an embedding layer to obtain $H_r$. 
    In this work, we include a total of 1824 categories (THING-EEG:1573+200, ImageNet-EEG:40, TUEV:6, TUAB:2, SEED: 3), and set k as 420 to ensure that the context window of the model is not exceeded.

    \subsubsection{RAG Example}
    In this study, when RAG is set to enabled, text in the following format will be inserted into the user's prompt:

    \begin{tcolorbox}[
        boxrule=1pt,
        arc=1mm,
        width=\columnwidth,
        title={RAG Prompt Example},
        colback=Emerald!10,colframe=cyan!40!black
    ]

    Following is the Feature Database Search Result, you can consider, but it may be wrong: \\
    If task is Emotion Recognition: positive \\
    If task is Abnormal Classification: abnormal \\
    If task is Event Classification: generalized periodic epileptiform discharges (GPED) \\
    If task is Visual Stimuli: bagel,omelet,card,pasta,pistachio,peppermint,casserole \textcolor{orange}{(Ignore 409 class..)}

    \end{tcolorbox}

    \subsubsection{RAG Pseudo code}
    \noindent
    \begin{minipage}{0.48\textwidth}
        \begin{algorithm}[H]
        \caption{Pre-compute and Store Supervision Feature}
        \label{alg:storage}
        \textbf{Input}: 
        $[\mathbf{X}_I, \mathbf{Y}_I, \mathcal{D}_I]$, 
        $[\mathbf{X}_T, \mathbf{Y}_T, \mathcal{D}_T]$, 
        $\text{DB}$
        
        \textbf{Parameters}: None
        
        \textbf{Output}: Updated $\text{DB}$
        
        \begin{algorithmic}
        \STATE \textbf{Modality Encoding}:
        \begin{align*}
        \mathbf{Z}_I &= \mathcal{F}_{\text{ViT}}(\mathbf{X}_I) \\
        \mathbf{Z}_T &= \mathcal{F}_{\text{BERT}}(\mathbf{X}_T)
        \end{align*}
        
        \STATE \textbf{Feature Normalization}:
        \begin{align*}
        \hat{\mathbf{z}}_I &= \frac{\mathbf{Z}_I}{\|\mathbf{Z}_I\|_2} \quad \text{(Image features)} \\
        \hat{\mathbf{z}}_T &= \frac{\mathbf{Z}_T}{\|\mathbf{Z}_T\|_2} \quad \text{(Text features)}
        \end{align*}
        
        \STATE \textbf{Store Features with Dataset Tags}:
        \FOR{each modality group $g$ in $\{\text{vis}, \text{txt}\}$}
            \IF {$g = \text{vis}$}
                \STATE $\hat{\mathbf{Z}} \gets \hat{\mathbf{z}}_I$, $\mathbf{Y} \gets \mathbf{Y}_I$, $\mathcal{D} \gets \mathcal{D}_I$
            \ELSE
                \STATE $\hat{\mathbf{Z}} \gets \hat{\mathbf{z}}_T$, $\mathbf{Y} \gets \mathbf{Y}_T$, $\mathcal{D} \gets \mathcal{D}_T$
            \ENDIF
            \STATE $\text{DB}.\text{extend}\left(\hat{\mathbf{Z}}, \mathbf{Y}, \mathcal{D}\right)$
        \ENDFOR
        
        \STATE \textbf{return} $\text{DB}$
        \end{algorithmic}
        \end{algorithm}
    \end{minipage}
    \hfill
    \begin{minipage}{0.48\textwidth}
        \begin{algorithm}[H]
            \caption{Query K Most-Relavant Supervision Feature}
            \label{alg:match_query}
            
            \textbf{Input}: $\text{DB}$, $\mathbf{X}_e$ \\
            \textbf{Parameter}: $K$ \\
            \textbf{Output}: Predicted labels $Y = [y_1, y_2, \dots, y_K]$
            
            \begin{algorithmic}[1]
            \STATE \textbf{Encode and normalize EEG sample:}
            
            $\mathbf{X}'_e = \mathcal{F}_{\theta}(\mathbf{X}_e), \quad \hat{\mathbf{v}} = \frac{\mathbf{X}'_e}{\|\mathbf{X}'_e\|_2}$
            
            \STATE \textbf{Group DB by dataset:}  
            
            $\mathcal{G} = \text{group\_by}(\text{DB}, \text{dataset})$
            
            \STATE \textbf{Initialize selected set:}  
            
            $\mathcal{S} = \emptyset$
            \STATE \textbf{Ensure at least one feature for each dataset}
            \FOR{each group $\mathcal{G}_i \in \mathcal{G}$}
                \STATE Compute similarities: $\mathcal{S}_i = [\hat{\mathbf{v}}^\top \mathbf{z}_j \mid \mathbf{z}_j \in \mathcal{G}_i]$  
                \STATE Select top-1: $\mathbf{z}^*_i = \arg\max \mathcal{S}_i$  
                \STATE Update $\mathcal{S} \gets \mathcal{S} \cup \{\mathbf{z}^*_i\}$
            \ENDFOR
            \STATE \textbf{Compute other simility:}
            \IF{$K > |\mathcal{G}|$}
                \STATE $\mathcal{C} = \text{DB} \setminus \mathcal{S}$  
                \STATE $\mathcal{S}_{\text{global}} = [\hat{\mathbf{v}}^\top \mathbf{z}_j \mid \mathbf{z}_j \in \mathcal{C}]$  
                \STATE $\mathcal{S}_{\text{extra}} = \text{top\_k}(\mathcal{C}, \mathcal{S}_{\text{global}}, K - |\mathcal{G}|)$  
                \STATE $\mathcal{S} \gets \mathcal{S} \cup \mathcal{S}_{\text{extra}}$
            \ENDIF
            
            \STATE \textbf{Return labels} 
            $[(\mathbf{z}) \mid \mathbf{z} \in \mathcal{S}]$
            \end{algorithmic}
        \end{algorithm}
    \end{minipage}

\section{Data Preprocessing Pipeline Detail}
    \label{sec:data_preprocerssing}
    The standardized preprocessing pipeline was applied uniformly to all EEG data, with each step implemented as follows:
    
    \textbf{Montage Unification}: A predefined set of 32 standard EEG channels was adopted (\texttt{'Fp1', 'Fp2', 'F7', 'F3', 'Fz', 'F4', 'F8', 'FC5', 'FC1', 'FC2', 'FC6', 'T7', 'C3', 'Cz', 'C4', 'T8', 'CP5', 'CP1', 'CP2', 'CP6', 'P7', 'P3', 'Pz', 'P4', 'P8', 'POz', 'O1', 'Oz', 'O2', 'AFz', 'CPz', 'FCz'}). To address channel variations across different datasets, the following procedures were implemented: First, valid EEG channels were automatically identified and retained. For missing standard channels, linear interpolation was employed for data reconstruction based on spatial distributions of neighboring electrodes, ensuring consistent channel space representation across all subjects. Finally, all channels were sorted in a standardized order to eliminate inconsistencies caused by variations in acquisition systems.
    
    \textbf{Resampling}: All EEG signals were uniformly resampled to 512 Hz using linear interpolation.
    
    \textbf{Fixed-Duration}: All EEG data were examined to ensure a consistent duration of 1 second per segment. For data shorter than 1 second, temporal repetition (repeat) was applied to extend the duration. For data exceeding 1 second, segmentation was performed to generate multiple 1-second epochs.
    
    
    

    

\section{Dataset Detail}

The details of each dataset with configuration are as below:

\textbf{THINGS-EEG} \cite{giffordLargeRichEEG2021,grootswagersHumanEEGRecordings2022}: Containing data from 10 human subjects performing an orthogonal target detection task under a rapid serial visual presentation (RSVP) paradigm. During the experiment, researchers presented subjects with a series of images while recording their EEG responses. Raw data were collected using a 64-channel system at a sampling rate of 1000 Hz. The dataset includes 16,540 training image conditions (each repeated 4 times) and 200 test image conditions (each repeated 80 times), totaling 82,160 image trials. We use a standard pipeline to process this dataset.

\textbf{ImageNet-EEG} \cite{palazzoGenerativeAdversarialNetworks2017,palazzoDecodingBrainRepresentations2021}: Containing EEG signals from 6 subjects viewing 40 object categories from the ImageNet dataset (50 images per category, totaling 2,000 images). The EEG data consists of 128-channel, 1kHz kHz-sampled 0.5-second EEG segments. We use file \textit{eeg\_signals\_raw\_with\_mean\_std.pth} to maintain full information. Then we repeat EEG segments in temporal scale to make 1 1-second sample. 


\textbf{SEED} \cite{zheng2015investigating,duan2013differential}: Contains EEG signals from 15 subjects watching 15 film clips (each about 4 minutes) designed to elicit positive, neutral, and negative emotions. Data were collected using a 62-channel ESI NeuroScan system at 1kHz sampling rate, downsampled to 200Hz with 0-75Hz bandpass filtering for artifact removal. We use a standard pipeline to process this dataset




\textbf{TUAB} \cite{obeidTempleUniversityHospital2016}: A subset of the TUH corpus includes normal and abnormal annotations from a clinical doctor. The dataset contains 22 channels in each segment and is collected at a sampling rate of 250Hz. Each segment contains 10 seconds of data, so we split it into 1-second samples. We only used a subset of the TUAB data in our study due to the large sample of the data. We use a standard pipeline to process this dataset.

\textbf{TUEV} \cite{obeidTempleUniversityHospital2016}: A subset of TUH corpus includes 16,986 annotated EEG segments from 10874 subjects classified into three signal classes and three noise classes: spike and sharp wave (SPSW), generalized periodic epileptiform discharges (GPED), periodic lateralized epileptiform discharges (PLED), eye movement (EYEM), artifact (ARTF), and background (BCKG). We use a standard pipeline to process this dataset.

\section{Instruction Construction Details}
    In this section, we demonstrate more details on instruction construction for Stage III. The data engineering details for Stage I and Stage III can be found in Appendix \ref{More Data Engineering Details}
    \subsection{Detail Data Statistic}
    \label{Detail Data Statistic}

    Based on the comprehensive data presented in Table \ref{tab:detail_static_appendix}, WaveMind Instruct constitutes a multi-modal EEG dataset comprising 338k instruction-response pairs curated from 5 data sources. The \textit{Consultation} scenarios are exclusively designed for TUEV and TUAB datasets with EEG-based event and abnormality detection, where Brain State neural patterns are analyzed. We add more 2k Reject Empty-EEG samples to enhance, avoid hallucinations without real EEG input, deliberately constructed by removing EEG inputs to counter hallucination vulnerabilities.

    \begin{table}[htbp]
            \centering
            \small
            \begin{tabularx}{\textwidth}{l l X c c c}
                \toprule
                \multirow{2}{*}{\textbf{Source Dataset}} & \multirow{2}{*}{\textbf{Task}} & \multirow{2}{*}{\textbf{Data Type}} & \multicolumn{2}{c}{\textbf{Scenario}} & \multirow{2}{*}{\textbf{\#Sample}} \\
                \cmidrule(lr){4-5}
                 &  &  & Consultation & Analysis & \\
                \midrule
                
                \multirow{4}{*}{THING-EEG} 
                    & \multirow{4}{*}{\makecell{Visual-Stimulus \\Interpretation}} 
                        & Closed-ended QA & \multirow{4}{*}{\ding{55}} & \multirow{4}{*}{\ding{51}} & 4,531 \\ 
                    && Open-ended QA & & & 18,128 \\ 
                    && Description & & & 54,303 \\ 
                    && MCQ & & & 76,962 \\ 
                \midrule
                   THING-EEG Total &  & & & & 153,924 \\ 
                \midrule
                
                \multirow{4}{*}{ImageNet-EEG} 
                    & \multirow{4}{*}{\makecell{Visual-Stimulus \\Interpretation}} 
                        & Closed-ended QA & \multirow{3}{*}{\ding{55}} & \multirow{4}{*}{\ding{51}} & 9,983 \\ 
                    && Open-ended QA & & & 44,936 \\ 
                    && Description & & & 28,596 \\ 
                    && MCQ & & & 19,628 \\ 
                \midrule
                   ImageNet-EEG Total &  & & & & 103,143 \\ 
                \midrule
                
                \multirow{3}{*}{TUEV} 
                    & \multirow{3}{*}{\makecell{Event \\Detection}} 
                        & Closed-ended QA & \multirow{3}{*}{\ding{51}} & \multirow{4}{*}{\ding{51}} & 2,241 \\ 
                    && Open-ended QA & & & 7,344 \\ 
                    && MCQ & & & 7,541 \\ 
                \midrule
                   TUEV Total &  & & & & 17,126 \\ 
                \midrule
                
                \multirow{3}{*}{TUAB} 
                    & \multirow{3}{*}{\makecell{Abnormality \\Detection}} 
                        & Closed-ended QA & \multirow{3}{*}{\ding{51}} & \multirow{4}{*}{\ding{51}} & 6,609 \\ 
                    && Open-ended QA & & & 19,828 \\ 
                    && MCQ & & & 12,760 \\ 
                \midrule
                   TUAB Total &  & & & & 39,197 \\ 
                \midrule

                \multirow{3}{*}{SEED} 
                    & \multirow{3}{*}{\makecell{Emotion \\Recognition}} 
                        & Closed-ended QA & \multirow{4}{*}{\ding{55}} & \multirow{4}{*}{\ding{51}} & 4,507 \\ 
                    && Open-ended QA & & & 13,524 \\ 
                    && MCQ & & & 10,270 \\ 
                \midrule
                    SEED Total & & & & & 28,301 \\


                \midrule
                    Empty Instruction & Reject Empty-EEG & & & & 1,676 \\

                \midrule
                    \textbf{Total} & & & &  & 338,952 \\

                \bottomrule
            
            \end{tabularx}
            \caption{Details of WaveMind-Instruct}
            \label{tab:detail_static_appendix}
        \end{table}

    \subsection{Quality Control Detail}
    \label{sec:human_eval}

    To mitigate model hallucination and enhance instructional diversity, a comprehensive three-stage quality control pipeline was implemented throughout the synthesis process. The errors in synthesized instructions may originate from the illusion of GPT, therefore, we use image captioning and factual information to guide the synthesis of cognitive and Brain State instructions, respectively. The input material used for guidance is important, necessitating rigorous pre-synthesis validation. We constructed a quality marking platform using \textit{Gradio} for manual inspection, where human validators are asked to conduct binary evaluations. Annotators verified the image caption and factual information separately. Before marking factual information, they were required to learn neuroscience knowledge in advance. As shown in Table \ref{tab:human_eval}, input material yielded exceptionally high fidelity scores of 0.945±0.023 for image captions and 0.966±0.013 for clinical annotations.

    \begin{table}[H]
        \centering
        \begin{tabular}{ccc}
        \toprule
             & \textbf{Image Caption} & \textbf{Clinical Annotation}\\
        \midrule
            Evaluator 1 & 0.975 & 0.983\\
            Evaluator 2 & 0.920 & 0.950 \\
            Evaluator 3 & 0.940 & 0.966 \\
        \midrule
            \rowcolor{lightblue} Total & 0.945±0.023 & 0.966±0.013 \\
        \bottomrule
        \end{tabular}
        \caption{\textbf{Human Evaluation Result}}
        \label{tab:human_eval}
    \end{table}
    
    During the in-synthesis phase, we explicitly prohibited the generation of instructions containing EEG-specific details or neural mechanism speculation through prompt engineering. In the subsequent post-synthesis phase, we implemented a deduplication process using a 0.75 similarity threshold that combined both 2-gram token overlap and ROUGE-L semantic equivalence metrics. This distillation pipeline preserved 70\% of the highest-quality outputs while effectively removing redundant content.

    \subsection{Synthesis Process and Prompt}
    \label{sec:synthesis_process}

        \subsubsection{Description Instruction}

        Description Instruction is only used in image supervision data. Firstly, we use the prompt below to ask the \textit{Qwen2.5-VL} to generate the caption of the image.
    
        \begin{tcolorbox}[
            boxrule=1pt,
            arc=1mm,
            width=\columnwidth,
            title={Prompt for Image Caption Generation},
            colback=SeaGreen!10!CornflowerBlue!10,colframe=RoyalPurple!55!Aquamarine!100!
        ]

        You are an AI visual assistant, and you are seeing a single image.\\
        Please describe this image
        
        \end{tcolorbox}

        Then, we use regular expressions to uniformly replace all vocabulary related to the image (e.g., "image", “picture”, etc.) with the "EEG" keyword. Subsequently, we select a question from the Question Set and match it with the processed description to generate clean data for feature alignment.

        \begin{tcolorbox}[
            boxrule=1pt,
            arc=1mm,
            width=\columnwidth,
            title={Question Set for Description Instruction},
            colback=SeaGreen!10!CornflowerBlue!10,colframe=RoyalPurple!55!Aquamarine!100!
        ]
    
             \begin{itemize}
                \item What is being viewed according to this EEG?
                \item Can the EEG data reveal what is being observed?
                \item What does the EEG suggest about the visual stimulus?
                \item What does the EEG suggest about the visual stimulus?
                \item From the EEG patterns, what could be the subject of observation?
                \item What visual scene does the EEG indicate?
                \item What object or event does the EEG imply is being watched?
                \item What might the EEG tell us about the current visual input?
                \item What scenario does the EEG suggest is being viewed?
                \item What is the likely visual focus inferred from the EEG?
            \end{itemize}
        
        \end{tcolorbox}

        \subsubsection{QA Instruction}

         \textbullet{   \textbf{Cognitive QA Synthesis}}

        Firstly, we use the following prompt to generate 5 Image Captions for \textit{Qwen2.5-VL}:
        
        \begin{tcolorbox}[
            boxrule=1pt,
            arc=1mm,
            width=\columnwidth,
            title={Prompt for Image Caption Generation},
            colback=OliveGreen!10,colframe=Green!70
        ]
    
        You are an AI visual assistant, and you are seeing a single image.\\
        Now you are asking to depict the image in detail, give me 10 captions for the image.\\
        You can describe the object, the color, the shape, the size, the background, and the location of the object in the image.\\
        Only provide the description of the image, do not provide any other information.\\

        As we know this object is: \textcolor{blue}{\{obj\_class\}}.\\
        
        Output format:\\
        Caption: [Description of the image]\\
        Caption: [Description of the image]\\
        Caption: [Description of the image]\\
        Caption: [Description of the image]\\
        Caption: [Description of the image]\\
        
        \end{tcolorbox}

        For each EEG sample, we extract two captions from its paired images and ask \textit{Qwen2.5-Instruct} to rewrite them accordingly:

        \begin{tcolorbox}[
            boxrule=1pt,
            arc=1mm,
            width=\columnwidth,
            title={Prompt for cognitive QA Instruction Synthesis},
            colback=OliveGreen!10,colframe=Green!70, breakable
        ]
    
        You are a helpful AI assistant. You need to generate a 10 single-round conversations.\\
        This is a visual reproduction task in an EEG neuroscience experiment. During the experiment, the subjects were watching images or other visual stimuli, and the electroencephalogram was recorded at this time.\\
        When constructing a construction, the role "humans" should ask specific questions related to what this subject/patient is watching.\\
        
        (1) Do not include any object name in the question. It is very important. For example, "what is the subject observing about the tree in the image?" is not allow, because the phase "about the tree" include the "tree" keyword, which is strongly prohibited. Instead, you can ask "what is this subject observing about the object in the image?".\\
        (2) The question should be easy to answer based on the image that the subject is watching. Should not include the complex reasoning.\\
        (3) Both question and answer should imply in text that the EEG is reflating the visual stimuli(image) that someone is watching. It is important to mention that the EEG is reflecting the visual stimuli that the subject is watching.\\
        (4) The word "EEG" or "electroencephalogram" or similar word MUST be mentioned in each conversation in either one of question and answer or both. The EEG is a tech that records the electrical activity of the brain.\\
        (5) Question and answer should all related to the given caption below. Answer should be detail and specific and only related to caption. Diverse sentence structures.\\
        (6) You may ask what is inside, the content, the color, the background,etc. But make sure it is easy to answer based on the image that the subject is watching.\\
        (7) 10 single-round conversations should be generated. They are isolated to each other.\\
        
        You will be given some caption with relate to the image which subjects is watching, and you need to generate a conversation based on the caption.\\

        Example:\\
        Question: According to the EEG, what content is the subject watching?\\
        Answer: I guess in this EEG segment that this human is watching a person holding a guitar. The guitar is black in color, and the person is wearing a red shirt.\\
        Question: How many objects are there in the image that this subject is watching in this EEG?
        Answer: In this EEG, The subject is watching two objects in the image. One object is a guitar, and the other object is a cup.\\

        Format:\\
        Question:xxx\\
        Answer:xxx\\
        Question:xxx\\
        Answer:xxx\\
        Question:xxx\\
        Answer:xxx\\

        Given caption with the image that subject is watching:\\
        \textcolor{blue}{\{Random Caption\}\\}
    
        \end{tcolorbox}

        \textbullet{\textbf{Brain State QA Instruction Synthesis}}

        We use the following prompt to make \textit{Qwen2.5-instruct} synthesize instructions, where Original QA is factual information of each annotation, which is manually refined and inspected.
        
        \begin{tcolorbox}[
            boxrule=1pt,
            arc=1mm,
            width=\columnwidth,
            title={Prompt Template for Instruction Rewrite},
            colback=OliveGreen!10,colframe=Green!70
        ]
        You are an artificial intelligence assistant, please help me Rewrite the following QA sentence.

        \#\#\# Requirement\\
        1. Your response must include questions and answers, using the following output format.\\
        2. Your sentence structure or tone should vary as much as possible. You can unleash your imagination and keep sentence long enough.\\
        3. Questions and answers should not involve detailed EEG features, as young doctors do not need external knowledge to answer them, so a macro perspective is necessary.\\
        4. The answer does not need to involve a reason. The answer needs to involve: determine the EEG state and guess what happen in this EEG. Doubtful or firm tone can be used. You can add other content if necessary.\\
        
        5. However, the keyword "is" should be avoided or changed. Structure like "xxx is xxx" should be change. again, sentence structure and tone should be very.\\
        6. Please, try your best change the noun to synonyms except event keyword. Diversity is very important! You can also use personification, such as "I think...", "Maybe...","I" as keyword is encouraged to use, but don't use so much.

        \textcolor{blue}{\{Task-Specific Requriement\}}

        \textcolor{blue}{\{Scenario Requirement\}}
        
        \#\#\# Origin QA\\
        Question:\textcolor{blue}{\{Question\}\\}
        Answer:\textcolor{blue}{\{Answer\}}\\

        \#\#\# Output Format\\
        Strictly follow the JSON structure below. \\
        “‘json
        {{ \\
        "QA": [ \\
        {{"Question": "..."}}, \\
        {{"Answer":"..."}}\\
        ] \\
        }} 
        “‘
        \end{tcolorbox}

        \begin{tcolorbox}[
            boxrule=1pt,
            arc=1mm,
            width=\columnwidth,
            title={Task-Specific Requirement for abnormality detection},
            colback=OliveGreen!10,colframe=Green!70
        ]
         7. Possible event are: spike and sharp wave (SPSW), generalized periodic epileptiform discharges (GPED), periodic lateralized epileptiform discharges (PLED), eye movement (EYEM), artifact (ARTF), background (BCKG). They are the only keyword that should not change, any other noun or condition can alse be appropriately changed.
         
        \end{tcolorbox}

        \begin{tcolorbox}[
            boxrule=1pt,
            arc=1mm,
            width=\columnwidth,
            title={Task-Specific Requriement for abnormality detection},
            colback=OliveGreen!10,colframe=Green!70
        ]
         7. Possible state are: abnormal, normal. They are the only keyword that should not change, any other noun or sentence expression can alse be appropriately changed.
         
        \end{tcolorbox}
    
        \begin{tcolorbox}[
            boxrule=1pt,
            arc=1mm,
            width=\columnwidth,
            title={Task-Specific Requriement for Emotion Recognition},
            colback=OliveGreen!10,colframe=Green!70
        ]
         7. Possible emotional are: positive, neutral, negative. They are the only keyword that should not change, any other noun or condition can alse be appropriately changed. For example, emotion is encouraged to change to other word.
         
        \end{tcolorbox}

        \textbullet{   \textbf{Scenario Requirement in Prompt Synthesis}}
        \label{Scenario Requirement in Rewrite Prompt}

         \begin{tcolorbox}[
            boxrule=1pt,
            arc=1mm,
            width=\columnwidth,
            title={Scenario Requirement},
            colback=OliveGreen!10,colframe=Green!70
        ]
         
         \begin{itemize}
           \item \textbf{Analysis} More daily and general conversations, where the identities of both parties are not regulated.
           \item \textbf{Consultation} A dialogue scenario that leans towards healthcare. The questioner is the patient, and the answerer is the doctor. It sometimes happens in occurred in the consultation room. The chat should be more colloquial. Remember, doctor should never explain the details in EEG, instead, focus on explaining on explain this disease (phenomenon) itself.
       \end{itemize}

        \end{tcolorbox}

    \subsection{Instruction Example}

        \mbox{}\\

         \begin{tcolorbox}[
            boxrule=1pt,
            arc=1mm,
            width=\columnwidth,
            title={Description Instruction Example},
            colback=Salmon!20, colframe=Salmon!90!Black
        ]
    
       \begin{itemize}
           \item \textbf{Question:} Can the EEG data reveal what is being observed?
           \item \textbf{Answer:} The EEG shows an Alsatian dog standing on a wooden deck, looking down at a rolled-up piece of paper on a green mat.
       \end{itemize}
        
        \end{tcolorbox}

         \begin{tcolorbox}[
            boxrule=1pt,
            arc=1mm,
            width=\columnwidth,
            title={QA Instruction Example},
            colback=Salmon!20, colframe=Salmon!90!Black
        ]
    
        \begin{itemize}
           \item \textbf{Question:} Can you identify what's occurring based on the EEG reading? There seems to be a recurring pattern that isn't quite normal.
           \item \textbf{Answer:} After reviewing the waveform, I suspect it might be a periodic lateralized epileptiform discharge (PLED). These recurring waves often indicate a localized seizure, possibly stemming from one hemisphere of the brain.
       \end{itemize}
        
        \end{tcolorbox}

         \begin{tcolorbox}[
            boxrule=1pt,
            arc=1mm,
            width=\columnwidth,
            title={MCQ Instruction Example},
            colback=Salmon!20, colframe=Salmon!90!Black
        ]
    
       \begin{itemize}
           \item \textbf{Question:} Could you confirm the brainwave patterns are within normal limits? Your task is determine if this EEG is abnormal or not. Please choose one option in the following. Only capital letter, no need to explain. (A)Normal (B)Abnormal
           \item \textbf{Answer:} A
       \end{itemize}
        
        \end{tcolorbox}

\section{More Data Engineering Details}
    \label{sec:More Data Engineering Details}
    \subsection{Encoder Representation Alignment}
    \label{sec:Hand-written_Description_appendix}
    We fully listed the correspondence between annotations and manual descriptions in Table~\ref{tab:hand_description}. Since contrastive learning requires fine-grained text representations, we endeavor to provide detailed manual descriptions to enhance the effectiveness of the alignment training.
    
    \begin{table}[htbp]
        \small
        \centering
        \renewcommand{\arraystretch}{1.2}
        \begin{tabularx}{1.05\columnwidth}{ m{1.2cm} c X }
        \toprule
        \textbf{Dataset} & \textbf{Annotation} & \textbf{Meanual Description} \\
        \midrule
        \multirow{6}{*}{TUEV} 
            & SPSW & SPSW (Spike and Sharp Wave) is abrupt, high-amplitude transients with epileptiform morphology, typically localized to focal regions. It has sharp rising phases and asymmetric waveforms. \\ 
            & GPED & GPED (Generalized Periodic Epileptiform Discharges) exhibits repetitive, bilateral synchronous spikes/sharp waves with uniform morphology. It has symmetric distribution. \\ 
            & PLED & PLED (Periodic Lateralized Epileptiform Discharges),  with unilateral or bilateral asynchronous periodic discharges, with "stereotyped" waveforms. It has lateralized localization. \\ 
            & EYEM & EYEM (Eye Movement), with low-frequency, high-amplitude frontal slow waves synchronized with blinks or saccades. It has rhythmicity, frontal dominance, and correlation with eye movement artifacts. \\ 
            & ARTF & ARTF (Artifact) with non-physiological signals (e.g., muscle noise, electrode pops) with abrupt amplitude shifts or high-frequency content. \\ 
            & BCKG & BCKG (Background): Stable, symmetric rhythms with age-appropriate amplitude. Absent epileptiform features, with preserved reactivity to stimuli. \\
        \midrule
        \multirow{2}{*}{TUAB} 
            & Abnormal & Abnormal EEG exhibits deviations such as slowed/fast rhythms, asymmetry, or with irregular amplitudes. May include spikes, sharp waves, or spike-wave complexes, or so on. \\
            & Normal & Normal EEG characterized by rhythmic patterns with stable amplitudes. The brain wave is normal and stable. \\
        \midrule
        \multirow{3}{*}{SEED} 
            & Positive & Positive EEG shows a joy, laughter, happiness, bright colors, celebration, warmth, atmosphere \\
            & Neutral & Neutral EEG shows a calm, peaceful, quiet, everyday activity, serene landscape, and a peaceful atmosphere \\ 
            & Negative & Negative EEG shows a sadness, loneliness, tears, grief, sorrowful atmosphere\\
        \bottomrule
        \end{tabularx}
        \caption{Hand-Written Description for Dual Modilities Label Representation Alignment}
        \label{tab:hand_description}
    \end{table}

    \subsection{WaveMind-Bench}
    \subsubsection{Testing Data Statistic}

    We constructed a total of 36k MCQs to test WaveMind, statistics can be found in Table \ref{tab:eeg_datasets}.

    \begin{table}[htbp]
        \centering
        \begin{tabularx}{0.8\textwidth}{lXll}
        \toprule
        \textbf{Source Dataset} & \textbf{Task} & \textbf{Options} & \textbf{\#Sample} \\
        \midrule

        \multirow{3}{*}{THING-EEG} & \multirow{3}{*}{Visual-Stimulus Interpretation} & 2 & 3,000 \\
         &  & 4 & 3,000 \\
         &  & 40 & 3,000 \\
         \midrule

        \multirow{3}{*}{ImageNet-EEG} & \multirow{3}{*}{Visual-Stimulus Interpretation} & 2 & 3,000 \\
         &  & 4 & 3,000 \\  
         &  & 40 & 3,000 \\ 
         \midrule

        \multirow{3}{*}{TUEV} & \multirow{3}{*}{Event Detection} & 2 & 3,000 \\
         &  & 4 & 3,000 \\  
         &  & 6 & 3,000 \\ 
         \midrule

        TUAB & abnormality detection & 2 & 3,000 \\
        \midrule

        \multirow{2}{*}{SEED} & \multirow{2}{*}{Emotion Recognition} & 2 & 3,000 \\
         &  & 3 & 3,000 \\

        \midrule
        \textbf{Total} &  &  & 36,000 \\

        \bottomrule
        \end{tabularx}
        \caption{Details of WaveMind-bench}
        \label{tab:eeg_datasets}
    \end{table}

    
    
    
        

    \subsubsection{Example}
    
        \begin{tcolorbox}[
                boxrule=1pt,
                arc=1mm,
                width=\columnwidth,
                title={WaveMind-Bench Example Question},
                colback=Emerald!10,colframe=cyan!40!black
            ]
           What is the event inside this EEG? Select one letter in the following. Do not explain. \\ (A)BCKG (B)ARTF
        \end{tcolorbox}

\section{Training Detail}
    
    \subsection{Training Setting}
\label{Training Setting Detail}
    We demonstrate the detailed training setting in Table \ref{tab:stage_1hyper} and Table \ref{tab:stage_23hyper}. In Stage I, we set the batch size to 1024 in order to balance the effectiveness and GPU memory. In Stage II and Stage III, we use Zero-2 as a parallel strategy to improve computational efficiency, which is different from Zero-3 in the original LLAVA setting.

    \begin{table}[htbp]
        \centering
        \begin{tabularx}{0.5\textwidth}{Xp{3cm}}
            \toprule
            \textbf{Hyperparameters} & \textbf{Value} \\
            \midrule
            Batch Size & 1024 \\
            Learning Rate & $1\times10^{-3}$ \\
            Optimizer & AdamW \\
            Warmup Ratio & 0.03 \\
            Epoch & 30 \\
            Early Stop & True \\
            Gradient checkpoint & False \\
            \bottomrule
            
        \end{tabularx}
        \caption{Training hyperparameters of Stage}
        \label{tab:stage_1hyper}
    \end{table}

    \begin{table}[htbp]
        \centering
        \begin{tabularx}{0.5\textwidth}{Xp{3cm}} 
            \toprule 
            \textbf{Hyperparameters} & \textbf{Value} \\
            \midrule 
            Batch Size & 32 (8 per device) \\
            Parallel strategy & Zero-2 \\
            LoRA Learning Rate & $1\times10^{-5}$ \\
            Modality Adapter Learning Rate & $3\times10^{-5}$ \\
            Optimizer & AdamW \\
            Warmup Ratio & 0.03 \\
            Epoch & 1 \\
            LORA & Unfreeze \\
            Modality Adapter & Unfreeze \\
            ATMM & Freeze \\
            Gradient checkpoint & True \\
            \bottomrule 
        \end{tabularx}
        \caption{Training hyperparameters of Stage I and Stage II}
        \label{tab:stage_23hyper}
    \end{table}

    \subsection{Data Augmentation}

    \paragraph{Global Z-score Normalization}
    For each EEG sample $X_e \in \mathbb{R}^{C \times T}$, where $C$ denotes channels and $T$ denotes time points, global z-score normalization computes a single mean and standard deviation across all elements of the sample. The global mean $\mu = \frac{1}{CT} \sum_{c=1}^{C} \sum_{t=1}^{T} X_e[c,t]$ and global standard deviation $\sigma = \sqrt{\frac{1}{CT} \sum_{c=1}^{C} \sum_{t=1}^{T} (X_e[c,t] - \mu)^2}$ are calculated, with $\sigma$ clamped to $\max(\sigma, 10^{-8})$ for numerical stability. The normalized output $X_e'$ is then:
    \[
    X_e' = \frac{X_e - \mu}{\max(\sigma, 10^{-8})}
    \]
    
    \paragraph{Global Standard Deviation Scaling}
    For each EEG sample $X_e$, this method scales the data by its global standard deviation without mean subtraction. The global standard deviation $\sigma = \sqrt{\frac{1}{CT} \sum_{c=1}^{C} \sum_{t=1}^{T} (X_e[c,t] - \mu)^2}$ (where $\mu$ is the global mean) is computed and clamped to $\max(\sigma, 10^{-8})$. The scaled sample $X_e'$ preserves the original mean while standardizing the variance:
    \[
    X_e' = \frac{X_e}{\max(\sigma, 10^{-8})}
    \]
    
    \paragraph{Channel-wise Z-score Normalization}
    For each EEG sample $X_e$, channel-wise z-score normalization independently processes every channel along the temporal axis. For channel $c$, the mean $\mu_c = \frac{1}{T} \sum_{t=1}^{T} X_e[c,t]$ and standard deviation $\sigma_c = \sqrt{\frac{1}{T} \sum_{t=1}^{T} (X_e[c,t] - \mu_c)^2}$ are computed, with $\sigma_c$ set to $10^{-8}$ if zero. The normalized output $X_e'$ is:
    \[
    X_e'[c, :] = \frac{X_e[c, :] - \mu_c}{\max(\sigma_c, 10^{-8})} \quad \text{for each channel } c
    \]
    
    \paragraph{Channel-wise Standard Deviation Scaling}
    For each EEG sample $X_e$, this method scales every channel by its channel-specific standard deviation without mean subtraction. For channel $c$, the standard deviation $\sigma_c = \sqrt{\frac{1}{T} \sum_{t=1}^{T} (X_e[c,t] - \mu_c)^2}$ (with $\mu_c$ as the channel mean) is computed and clamped to $\max(\sigma_c, 10^{-8})$. The scaled sample $X_e'$ maintains the original channel means:
    \[
    X_e'[c, :] = \frac{X_e[c, :]}{\max(\sigma_c, 10^{-8})} \quad \text{for each channel } c
    \]
    \paragraph{Amplitude Fluctuation Augmentation}
    For each EEG sample $X_e \in \mathbb{R}^{C \times T}$, this augmentation method applies random amplitude scaling across all channels and time points. A scaling factor $s_e$ is independently sampled for each sample from a uniform distribution $\mathcal{U}(1 - a, 1 + a)$, where $a$ denotes the fluctuation amplitude (e.g., $a=0.1$ for $\pm 10\%$ variation). The scaled output $X_e'$ preserves the temporal structure while introducing amplitude variability:
    \[
    X_e' = s_e \cdot X_e, \quad \text{where} \quad s_e \sim \mathcal{U}(1 - a, 1 + a)
    \]
    This operation maintains the relative channel relationships and temporal dynamics while simulating natural amplitude variations observed in EEG recordings. Each sample undergoes independent scaling, ensuring batch-level diversity in the augmented dataset.

    \clearpage
    
    \subsection{Dual-Supervision Representation Alignment}

    \subsubsection{Pseudo code for Representation Alignment}

   \begin{algorithm}[h]
        \caption{Dual Modalities Loss Computation}
        \label{alg:clip_loss}
        \textbf{Input}: 
        $\mathbf{X}_e$, 
        $\mathbf{X}_I$,
        $\mathbf{X}_T$
        
        \textbf{Parameters}: 
        $\lambda \in [0,1]$, logit scale $\tau$
        
        \textbf{Output}: Loss $\mathcal{L} \in \mathbb{R}$
        
        \begin{algorithmic}[1]
        \STATE Encode modalities:
        
            $\mathbf{X}'_e = \mathcal{F}_{\theta}(\mathbf{X}_e)$,
            $\mathbf{Z}_I = \mathcal{F}_{ViT}(\mathbf{X}_I)$,
            $\mathbf{Z}_T = \mathcal{F}_{BERT}(\mathbf{X}_T)$,
        \STATE Normalize features: 
        
        $\hat{\mathbf{e}} = \frac{\mathbf{X}'_e}{\|\mathbf{X}'_e\|_2}$,
        $\hat{\mathbf{z}}_I = \frac{\mathbf{Z}_I}{\|\mathbf{Z}_I\|_2}$,
        $\hat{\mathbf{z}}_T = \frac{\mathbf{Z}_T}{\|\mathbf{Z}_T\|_2}$
        
        \STATE Split samples by modalities:
        
        - $\mathcal{D}_{\text{vis}} = \{i | \text{Label\_Type}_i = \text{Image}\}$
            
        - $\mathcal{D}_{\text{txt}} = \{i | \text{Label\_Type}_i = \text{Text}\}$
        
        \STATE Compute losses by group:
        \FOR{\textbf{group} $g$ in $\{\text{vis}, \text{txt}\}$}
            \STATE Let $N_g = |\mathcal{D}_g|$
            \STATE Extract group features: 
                $\hat{\mathbf{E}}_g = [\hat{\mathbf{e}}^{(i)}]_{i\in\mathcal{D}_g}$,
                $\hat{\mathbf{Z}}_g = [\hat{\mathbf{z}}_g^{(i)}]_{i\in\mathcal{D}_g}$
            
            \STATE Compute InfoNCE loss:
            $$
            \mathcal{L}_g = -\frac{1}{N_g}\sum_{i=0}^{N_g-1} \log\frac{
                \exp(\tau \cdot \langle \hat{\mathbf{e}}^{(i)}, \hat{\mathbf{z}}_g^{(i)} \rangle)
            }{
                \sum_{j=0}^{N_g-1} \exp(\tau \cdot \langle \hat{\mathbf{e}}^{(i)}, \hat{\mathbf{z}}_g^{(j)} \rangle)
            }
            $$
        \ENDFOR
        
        \STATE Combine losses: $\mathcal{L} = \lambda \mathcal{L}_{\text{vis}} + (1-\lambda)\mathcal{L}_{\text{txt}}$
        \STATE \textbf{return} $\mathcal{L}$
        \end{algorithmic}
    \end{algorithm}

    \subsubsection{Confusion Matrix Analysis}

\begin{figure}[h!]
                \centering
                \includegraphics[width=\linewidth]{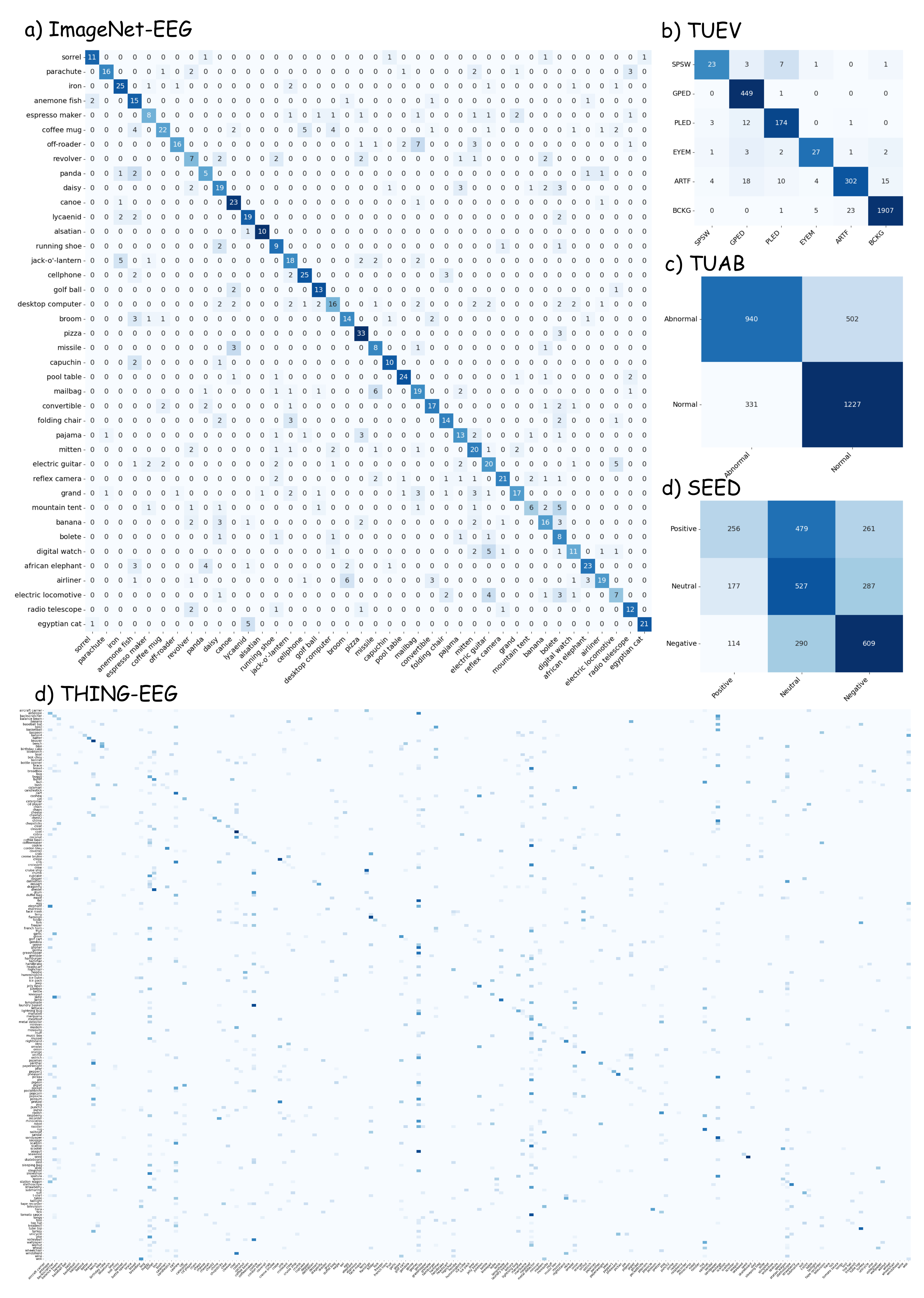}
                \caption{\textbf{Retrival confusion matrix of each dataset}. For THING-EEG, the zero-shot result under the SD protocol is visualized.}
                \label{fig:confusion-matrix}
             \end{figure}

        Figure \ref{fig:confusion-matrix} shows the retry confusion matrix analysis results of the ATMM model on various datasets.
         Each matrix is generated based on a retry prediction of 3000 random samples, where THING-EEG adopts a zero sample learning paradigm (using only the sub01-sub09 data of the subjects), and the remaining datasets are closed-set tasks that cover all subject data.
        
        It is worth noting that the confusion matrix of ImageNet EEG exhibits clearer main diagonal features (Figure 3a), which is consistent with the quantitative results in Table 1 and may be due to the advantages of this dataset in terms of experimental design standardization and data quality control.
         The confusion matrix of the SEED dataset (Figure 3d) reveals a significant phenomenon of category confusion: neutral emotion samples are heavily misclassified into positive and negative categories, suggesting that there may be feature expression ambiguity in this emotion dimension.

    \subsubsection{Representation Analysis}

\begin{figure}[h!]
            \centering
            \includegraphics[width=1\linewidth]{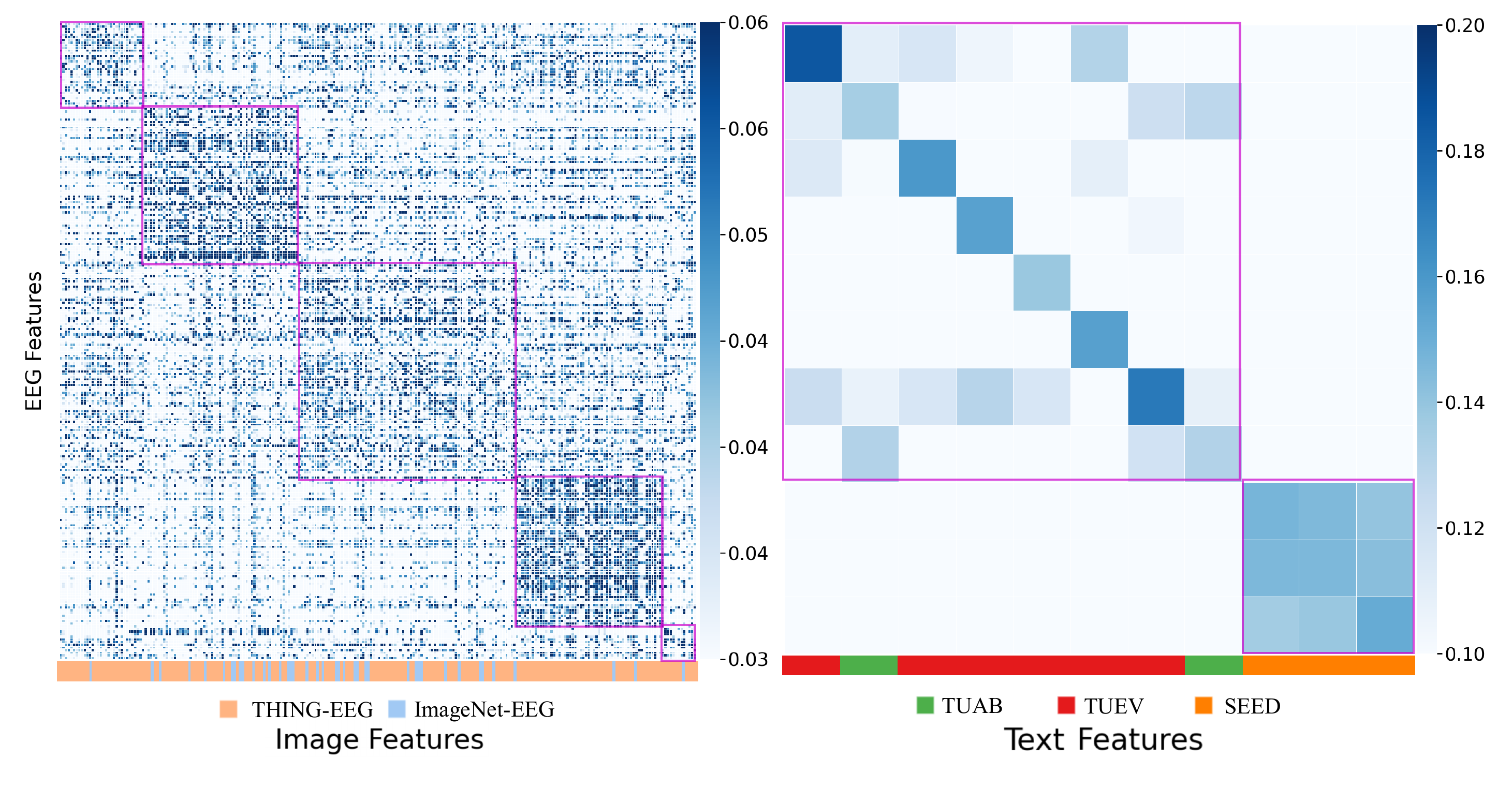}
            \caption{\textbf{Visualization of Representation Similarity Matrix (RSM).} 
                \protect\underline{Left} is RSM between EEG feature and image-label feature on Image-EEG dataset, while \protect\underline{Right} is RSM between EEG feature and text-label feature on Text-EEG dataset. The bottom color bar indicates the different datasets.
            }
            \label{fig:Rep_appendix}
        \end{figure}

        Figure \ref{fig:Rep_appendix} presents the analysis results of Representational Similarity Matrices (RSMs) used to evaluate the association between EEG features and embeddings from different modalities of labels. The left panel compares EEG features with image supervision features (Image-EEG data), while the right panel contrasts EEG features with text supervision features (Text-EEG data). Applying K-means clustering to both datasets reveals clear clustering phenomena. The RSM based on text supervision (Text-EEG) exhibits two main clusters: samples from the SEED dataset, which describe emotions, form a distinct cluster and are semantically differentiated from samples in the TUAB and TUEV datasets, which describe characteristics. In contrast, the RSM based on image supervision is divided into five clusters according to semantic information. This indicates that the EEG features have effectively captured the semantic differences embedded in the corresponding labels.

        Further cross-dataset analysis shows that EEG features from different sources exhibit a mixed distribution in the embedding space rather than strict within-dataset clustering. This cross-domain feature interpenetration demonstrates that the multi-dataset training strategy has successfully aligned the representation spaces. Notably, the main diagonal regions of both RSM matrices display significantly high-intensity features, directly confirming that the model establishes stable semantic association mappings in the feature space. These results, which combine semantic sensitivity with cross-dataset robustness, provide crucial theoretical support for building a universal large-scale EEG model framework.

\section{Open-Source Asset}

We have uploaded our source code, synthesis instruction-tuning data, and proposed benchmark to the GitHub repository: \url{https://github.com/ZiyiTsang/WaveMind}. However, due to strict privacy restrictions governing the EEG dataset, we are unable to share processed EEG data. Instead, we have open-sourced the EEG processing code to the community to ensure reproducibility.

\begin{itemize}
    \item Project Page: \url{https://ziyitsang.github.io/WaveMind/}
    \item MLLM Source Code: \url{https://github.com/ZiyiTsang/WaveMind/tree/master/EEGLLM/LLaVA/llava}
    \item README:\url{https://github.com/ZiyiTsang/WaveMind/blob/master/README.md}
    \item Data Engineering: \url{https://github.com/ZiyiTsang/WaveMind/tree/master/Data_Engineering}
    \item EEG dataset and preprocessing: \url{https://github.com/ZiyiTsang/WaveMind/blob/master/data}
    \item WaveMind-Instruct-338K: Still preparing. We will release it in the Huggingface repo soon.
    \item WaveMind-Bench-12K:\url{https://huggingface.co/datasets/CocoNutZENG/WaveMind_Bench}
    \item Serving:\url{https://github.com/ZiyiTsang/WaveMind/blob/master/EEGLLM/Predict.py}
    \item Model Checkpoint: Still preparing. We will release it in the Huggingface repo after the paper is accepted.
\end{itemize}

\end{document}